\pdfoutput=1
\documentclass[journal,10pt]{IEEEtran}
\IEEEoverridecommandlockouts
\usepackage{amsmath,graphicx,algorithmic,algorithm,color,balance}
\usepackage{array}
\usepackage{amsfonts}
\usepackage{hyperref}
 \DeclareMathOperator*{\argmin}{argmin}

\newtheorem{definition}{\textbf{Definition}}

\newtheorem{lemma}{\textbf{Lemma}}

\newtheorem{proposition}{\textbf{Proposition}}

\renewcommand{\l}{\ell}
\newcommand{\norm}[1]{\lVert#1\rVert}
\newcommand{\card}[1]{\lvert#1\rvert}

\newcommand\ip[2]{\langle #1, #2\rangle}

\newcommand{\E}{{\mathcal{E}}}
\newcommand{\V}{{\mathcal{V}}}
\newcommand{\G}{{\mathcal{G}}}

\newcommand{\R}{{\mathbb{R}}}

\newcommand\opnorm[1]{|\!|\!| #1|\!|\!|}

\newcommand{\Rbb}{\mathbb{R}}

\renewcommand{\L}{{\mathcal{L}}}

\newcommand{\Scal}{\mathcal{S}}

\newcommand{\transpose}{{\!\scriptscriptstyle\mathrm T}}  % Transpose

\def\L{{\mathbf{\cal L}}}

\DeclareMathOperator*{\argmax}{arg\,max}

\begin{document}

\title{Distributed Signal Processing via Chebyshev Polynomial Approximation}
\author{
%    \IEEEauthorblockN{David I Shuman\IEEEauthorrefmark{1}, Pierre Vandergheynst\IEEEauthorrefmark{2}, Daniel Kressner\IEEEauthorrefmark{3}, Pascal Frossard\IEEEauthorrefmark{2}}
        \IEEEauthorblockN{David I Shuman, Pierre Vandergheynst, Daniel Kressner, Pascal Frossard}
  \thanks{David I Shuman is with the Department of Mathematics, Statistics, and Computer Science, Macalester College, St. Paul, MN 55105, USA (email: dshuman1@macalester.edu). Pierre Vandergheynst and Pascal Frossard are with the Signal Processing Laboratory, Ecole Polytechnique F{\'e}d{\'e}rale de Lausanne (EPFL), Institute of Electrical Engineering, CH-1015 Lausanne, Switzerland (email: \{pierre.vandergheynst,~pascal.frossard\}@epfl.ch). Daniel Kressner is with the Numerical Algorithms and High Performance Computing Laboratory, Ecole Polytechnique F{\'e}d{\'e}rale de Lausanne (EPFL), Institute of Electrical Engineering, CH-1015 Lausanne, Switzerland (email: daniel.kressner@epfl.ch).}
    \thanks{Part of the work reported here was presented at the \emph{IEEE International Conference on Distributed Computing in Sensor Systems (DCOSS), June 2011, Barcelona, Spain}.}
    \thanks{This work was supported in part by FET-Open grant number 255931 UNLocX and FNS grant number 200021-118230.}
    \thanks{The authors would also like to thank Jalal Fadili for his help deriving the bound in Proposition \ref{Prop:denoising}.}
    \thanks{MATLAB code for all numerical experiments in this paper is available at \url{http://www.macalester.edu/\textasciitilde dshuman1/publications.html}}
    }
    \maketitle
  %      \medskip 
%    \IEEEauthorblockA{\IEEEauthorrefmark{1}Macalester College, Department of Mathematics, Statistics, and Computer Science, St. Paul, MN 55105
%    \\Email: dshuman1@macalester.edu}
%    \medskip
%    
%    \IEEEauthorblockA{\IEEEauthorrefmark{2}Ecole Polytechnique F{\'e}d{\'e}rale de Lausanne (EPFL), Signal Processing Laboratory, CH-1015 Lausanne, Switzerland
%        \\Email: \{pierre.vandergheynst,~pascal.frossard\}@epfl.ch}
%        \medskip
%        
%    \IEEEauthorblockA{\IEEEauthorrefmark{3}Ecole Polytechnique F{\'e}d{\'e}rale de Lausanne (EPFL), Numerical Algorithms and High Performance Computing Laboratory, CH-1015 Lausanne, Switzerland
%    \\Email: daniel.kressner@epfl.ch}
%\ninept
%

%
\begin{abstract}
Unions of graph
multiplier operators are an important class of linear operators for processing signals defined on graphs. We present a novel method to efficiently distribute the application of these
operators. The proposed method
features approximations of the graph 
multipliers by shifted Chebyshev polynomials, whose recurrence relations make them readily amenable to distributed computation.
We demonstrate how the proposed method can be applied to distributed processing tasks such as smoothing, denoising, inverse filtering,
and semi-supervised classification, and 
show that the communication requirements of
the method scale gracefully with the size of the network.
\end{abstract}

\section{Introduction}\label{sec:intro}

In distributed signal processing tasks, the data to be processed is physically separated and cannot be transmitted to a central processing entity.
This separation may be due to engineering limitations such as the limited communication range of wireless sensor network nodes, privacy concerns, or
%engineering 
design considerations. Even when high-dimensional data can be processed centrally, %for example, 
it may be more efficient to process it with parallel computing. % on %a number of
%different
%multiple GPUs.
It is therefore important %critical
to develop distributed  %for %in-network
data processing algorithms that %help
balance the trade-offs between performance,
communication bandwidth, and computational complexity (speed).

\subsection{The Communication Network and Signals on the Network}

For concreteness, we focus throughout the paper on distributed processing examples in wireless sensor networks; however, the problems we consider could arise in a number of  different settings. %of the above situations.
Due to the limited communication range of wireless sensor nodes, % (resulting either from hardware constraints or from the desire to conserve energy),
 each sensor node in a large network is likely to communicate with only a small number of other nodes in the network. To model the communication patterns, we can write down a graph with each vertex corresponding to a sensor node and each edge corresponding to a pair of nodes that communicate. Moreover, because the communication graph is a function of the distances between nodes, it often captures spatial correlations between sensors' observations as well.
 That is, if two sensors are close enough to communicate, their observations are %more
 likely to be correlated.
We can further specify these %expected
spatial correlations by adding weights to the edges of the %communication
graph, with higher weights associated to edges connecting sensors with closely correlated observations.

We model the communication network with an undirected, weighted graph $\G = \{\V,\E,w\}$, which consists of a set of vertices $\V$, a
set of edges $\E$, and a weight function $w:\E\to\mathbb{R}^+$ that assigns a non-negative weight to each edge. We assume the number of nodes in the network, $N=|\V|$, is finite, and the graph is connected. The adjacency (or weight) matrix $\mathbf{W}$ for a weighted graph $\G$ is the $N\times N$ matrix with entries ${W}_{m,n}$, where
\begin{equation*}
{W}_{m,n} =
\begin{cases}
w(e), &\mbox{ if $e\in \E$ connects vertices $m$ and $n$} \\
0, &\mbox{ otherwise}
\end{cases}~.
\end{equation*}
Therefore, the weighted graph $\G$ can be equivalently represented as the triplet $\{\V,\E,\mathbf{W}\}$. The degree of each vertex is the sum of the weights of all the edges incident
to it. We define the degree matrix $\mathbf{D}$ to be the diagonal matrix with the $n$th diagonal entry ${D}_{n,n}$ equal to the sum of the entries in the $n^{\mathrm{th}}$ row of $\mathbf{W}$.

A signal or function $f: \V \rightarrow \Rbb$ defined on the vertices of the graph may be represented as a vector $\mathbf{f} \in \Rbb^N$, where the $n^{\mathrm{th}}$ component of the vector $\mathbf{f}$ represents the function value at the $n^{\mathrm{th}}$ vertex in $\V$. Throughout, we use bold font to denote matrices and vectors, and we denote the $n^{\mathrm{th}}$ component of a vector $\mathbf{f}$ by either $f(n)$ or $f_n$.

\subsection{Distributed Signal Processing Tasks} \label{Se:dist_tasks}

We consider sensor networks whose nodes can only send messages to their local neighbors
 (i.e., 
they cannot communicate directly with a central entity). Much
of the literature on
distributed signal processing in such settings (see, e.g., \cite{Rabbat}-\nocite{predd, olfati,dimakis}\cite{sandryhaila_consensus} and references therein) focuses %is focused
on %distributedly
coming to an agreement on simple features of the observed signal (e.g., consensus averaging, parameter estimation).
We are more interested in processing the full function in a distributed manner, with each node having its own objective.
Some example tasks under this umbrella include:
\begin{itemize}
\item \emph{Distributed denoising} -- In a sensor network of $N$ sensors, a noisy $N$-dimensional signal is observed, with each component of the signal corresponding to the observation at one sensor location. Using the prior knowledge that the denoised signal should be smooth or piecewise smooth with respect to the underlying weighted graph structure, the sensors' task is to denoise each of their components of the signal by iteratively passing messages to their local neighbors and performing computations.
\item \emph{Distributed semi-supervised learning / transductive classification} -- A class label is associated with each sensor node; however, only a small number of nodes in the network have knowledge of their labels. The cooperative task is for each node to learn its label by iteratively passing messages to its local neighbors and performing computations.
\end{itemize}

\subsection{Related Work}
The tasks mentioned in Section \ref{Se:dist_tasks} as well as other similar tasks have been considered recently in centralized settings in 
the 
fields of machine learning and signal processing on graphs \cite{shuman2013emerging}. For example,
\cite{smola}-\nocite{zhou_scholkopf}\cite{reg_discrete} consider general regularization frameworks on weighted graphs;
\cite{zhu_g}-\nocite{harmonic}\nocite{belkin_matveeva}\nocite{zhou_bousquet}\nocite{delalleau}\nocite{chapelle}\nocite{ando_zhang}\cite{johnson_zhang} present graph-based semi-supervised learning methods; and \cite{bougleux}-\nocite{elmoataz, hancock}\cite{peyre_nlr} consider regularization and filtering on weighted graphs for image and mesh processing.
Spectral regularization methods for ill-posed inverse problems (see, e.g., \cite{rosasco} and references therein) are also closely related.

Also in a centralized setting, \cite{LTS-ARTICLE-2009-053} shows that %the 
a truncated Chebyshev polynomial expansion efficiently approximates the application of a spectral graph wavelet transform.
The truncated Chebyshev polynomial expansion technique is originally introduced in \cite{druskin} 
in the context of approximately computing the product of a matrix function and a vector. 
In Section \ref{Se:background}, we discuss the connection between the graph  
multiplier operators we define and %the 
more general matrix functions.

In the distributed setting, reference \cite{wagner} considers denoising via wavelet processing and \cite{barbarossa} presents a denoising algorithm that projects the measured signal onto a low-dimensional subspace spanned by smooth functions.
References \cite{guestrin}-\nocite{predd_tit}\nocite{dlasso_conf}\cite{dlasso} consider different distributed regression problems. Reference \cite{thanou_quantized} extends the approach proposed in this paper by examining robustness to quantization noise. Segarra et al. \cite{segarra2015distributed,segarra_tsp} approximate general linear transformations by what we define in Section \ref{Se:matrix_functions} as graph multiplier operators. Infinite impulse response (IIR) graph spectral filters, which have  recently been introduced in  \cite{shi2015infinite,loukas2015distributed}, comprise another approach to many distributed graph signal processing tasks. These filters, which we discuss in more detail in Section \ref{Se:rational}, can be written as the ratio of two polynomial functions.

\subsection{Main Contributions}

In the the initial presentation of this work \cite{shuman_DCOSS_2011},
we extend the Chebyshev polynomial approximation method to the general class of unions of graph Fourier
multiplier operators, and show how the recurrence properties of the Chebyshev polynomials also enable distributed application of these operators. The communication requirements for distributed computation using this method scale gracefully with the number of sensors in the network (and, accordingly, the size of the signals). 

Our main contributions in this paper
are to i) generalize graph Fourier multiplier operators to %generalized 
graph multiplier operators (to be defined in detail in Section \ref{Se:background});
ii) show that the application of linear operators that are unions of
graph multiplier operators is a key component of distributed signal processing tasks such as  distributed smoothing, denoising, inverse filtering, and semi-supervised learning; %s \ref{Se:GFM} and \ref{Se:generalized}, respectively); 
iii) present a novel method to
efficiently distribute the application of the graph %Fourier
multiplier operators to %the
high-dimensional signals; iv) provide theoretical bounds on the approximation error incurred by the proposed method; and v) theoretically and numerically compare the proposed method to alternative distributed computation methods.

The remainder of the paper is as follows. In the next section, we provide %necessary
%notations and
some background from
spectral graph theory and matrix function theory, and introduce graph multiplier operators. 
In Section \ref{Se:applications}, we provide examples of distributed signal processing tasks that feature the application of graph multiplier operators. In Section \ref{Se:chebyshev}, we introduce a method to efficiently approximate these operators in a distributed setting via shifted Chebyshev polynomials. We discuss alternative methods to perform these approximate distributed computations in Section \ref{Se:other_methods}, and we theoretically and numerically compare these alternative methods. % in Section \ref{Se:comparison}. 
In Section \ref{Se:lasso}, we show how these methods can also be used to perform distributed wavelet denoising with the lasso regularization problem.
Section \ref{Se:conclusion} concludes the paper.

\section{Matrix Functions and Graph Multiplier Operators} \label{Se:background}
In this section, we leverage notation from the theory of matrix functions to introduce a class of operators that we call \emph{graph multiplier operators}. We also relate these operators to multiplier operators from classical Fourier analysis.

\subsection{Matrix Functions}\label{Se:matrix_functions}
Functions of matrices \cite{higham} appear throughout mathematics, science, and engineering. While functions of more general matrices can be defined via the Jordan canonical form (e.g., \cite[Definition 1.2, p. 3]{higham}), we restrict our attention in this paper to the simpler case of functions of real symmetric positive semi-definite matrices. Such a matrix $\mathbf{P} \in \Rbb^{N\times N}$ has a complete set of orthonormal eigenvectors $\left\{\mathbf{u}_{\l}\right\}_{\l=0,1,\ldots,N-1}$ and associated real, non-negative eigenvalues $\left\{\lambda_{\l}\right\}_{\l=0,1,\ldots,N-1}$ satisfying $\mathbf{P}\mathbf{u}_{\l}=\lambda_{\l}\mathbf{u}_{\l}$. That is, $\mathbf{P}$ admits a spectral decomposition
%\begin{align*}
$\mathbf{P}=\mathbf{U}\boldsymbol{\Lambda}\mathbf{U}^*$,
%\end{align*}
where $\mathbf{U}$ is the $N \times N$ matrix with the $(\l+1)^{\mathrm{th}}$ column equal to the eigenvector $\mathbf{u}_{\l}$, and $\boldsymbol{\Lambda}$ is the $N \times N$ diagonal matrix with the $(\l+1)^{\mathrm{th}}$ diagonal element equal to $\lambda_{\l}$. 
Without loss of generality, we assume the eigenvalues to be ordered as
\begin{align*}
0 \leq \lambda_0 \leq \lambda_1 \leq ... \leq \lambda_{N-1}:=\lambda_{\max}.
\end{align*}
Given a function $g(\cdot)$ well-defined on the spectrum $\sigma(\mathbf{P}):=\{\lambda_0, \lambda_1,\ldots,\lambda_{\max}\}$, the corresponding
\emph{matrix function} $g(\mathbf{P})$ is defined (e.g., \cite[p.3]{higham}) as
\begin{align}\label{Eq:mat_fun}
g(\mathbf{P}):=\mathbf{U}g(\boldsymbol{\Lambda})\mathbf{U}^*:=\mathbf{U}\left[
\begin{array}{ccc}
g(\lambda_0) &&\mathbf{0} \\
&\ddots & \\
\mathbf{0}&&g(\lambda_{N-1})
\end{array}
\right]\mathbf{U}^*.
\end{align}

The class of operators that can be written as matrix functions of $\mathbf{P}$ can be equivalently characterized as follows.
\begin{proposition} \label{Prop:char}
For a fixed real symmetric positive semi-definite matrix $\mathbf{P}$, the following are equivalent:
\begin{itemize}
\item[(a)] $\mathbf{\Psi}=g(\mathbf{P})$ for some $g: \sigma(\mathbf{P}) \rightarrow \Rbb$. %is a graph multiplier operator with respect to $\mathbf{P}$.
\item[(b)] $\mathbf{\Psi}$ and $\mathbf{P}$ are simultaneously diagonalizable by a unitary matrix; i.e., there exists a unitary matrix $\mathbf{U}$ such that $\mathbf{U}^{*}\mathbf{\Psi}\mathbf{U}$
and  $\mathbf{U}^{*}\mathbf{P}\mathbf{U}$ are both diagonal matrices.
\item[(c)] $\mathbf{\Psi}$ and $\mathbf{P}$ commute; i.e., $\mathbf{\Psi}\mathbf{P}=\mathbf{P}\mathbf{\Psi}$.
\end{itemize}
\end{proposition}
\begin{IEEEproof}[Proof of Proposition \ref{Prop:char}]  (a) implies (b) because of the definition \eqref{Eq:mat_fun} of $g(\mathbf{P})$, and (b) implies (a) if we set $g(\lambda_{\l})$ to the $(\l+1)^{\mathrm{th}}$ diagonal element of  $\mathbf{U}^{*}\mathbf{\Psi}\mathbf{U}$.
The equivalence between (b) and (c) is shown in \cite[Corollary 4.5.18]{horn}.
\end{IEEEproof}

\subsection{Graph Multiplier Operators}
In the context of distributed signal processing tasks, we are particularly interested in functions of symmetric matrices whose sparsity pattern is consistent with the communication structure of the network. 
\begin{definition}\label{Def:gmo}
$\mathbf{\Psi}$ is a %generalized 
\emph{graph multiplier operator} with respect to the graph $\G$ if there exists a real symmetric positive semi-definite matrix $\mathbf{P}$ and a function $g: \sigma(\mathbf{P}) \rightarrow \Rbb$ such that 
\begin{itemize}
\item[(i)] $\mathbf{\Psi}=g(\mathbf{P})=\sum_{\l=0}^{N-1} g(\lambda_{\l})\mathbf{u}_{\l}\mathbf{u}_{\l}^*$, and
\item[(ii)] $P_{i,j}=0$ if $W_{i,j}=0$ and $i\neq j$; i.e., $\mathbf{P}$ has the same sparsity pattern as the graph Laplacian $\L$ of the graph $\G$.
\end{itemize}
\end{definition}

In order for the distributed computational methods we introduce in Sections \ref{Se:chebyshev} and \ref{Se:other_methods} %of the next section
to be applicable to a wider range of applications, we can generalize slightly from graph multiplier operators to %Fourier multipliers to
\emph{unions of graph multiplier operators}. A union of graph multiplier operators is a linear operator
$\mathbf{\Phi}: \Rbb^N \rightarrow \Rbb^{\eta N}$ ($\eta \in \{1,2,\ldots\}$) 
that can be written as 
\begin{align}\label{Eq:union_def}
&~
\begin{array}{c}
{\color{blue} N } \\
{\color{blue} \overbrace{~\hspace{.2in}~}} 
\end{array}
\vspace{-3in} \nonumber \\
\mathbf{\Phi} = \left[
\begin{array}{c}
g_1(\mathbf{P}) \\
g_2(\mathbf{P}) \\
\vdots \\ 
g_{\eta}(\mathbf{P}) 
\end{array}
\right]=
&\left[
\begin{array}{c}
\mathbf{\Psi}_1 \\
\mathbf{\Psi}_2 \\
\vdots \\ 
\mathbf{\Psi}_{\eta} 
\end{array}
\right] 
{\color{blue}
\left.
\begin{array}{c}
~ \\
~ \\
~ \\ 
~ 
\end{array}
\hspace{-.2in}
\right\} {\eta N} }~~.
\end{align}
The application of the operator $\mathbf{\Phi}$ to a function $\mathbf{f}$ %is shown in Figure \ref{Fig:union}, and 
can equivalently be written as
\begin{align} \label{Eq:operator_def}
&\left(\mathbf{\Phi f}\right)_{(j-1)N+n} = \sum_{\l=0}^{N-1}g_j(\lambda_{\l}) %\hat{f}(\l) 
\langle \mathbf{f},\mathbf{u}_{\l}\rangle u_{\l}(n), \\
& \hspace{2.2cm}\hbox{for }j\in\{1,2,\ldots,\eta\},~ n\in\{1,2,\ldots,N\}. \nonumber
\end{align}

\subsection{Graph Fourier Multiplier Operators} \label{Se:gfmo}
When the matrix $\mathbf{P}$ in Definition \ref{Def:gmo} is the graph Laplacian $\L$, we call $\mathbf{\Psi}$ a \emph{graph Fourier multiplier operator}.
The non-normalized graph Laplacian is the real symmetric matrix $\L:=\mathbf{D}-\mathbf{W}$, the difference between the degree matrix and the weighted adjacency matrix (see, e.g., \cite{chung,spielman_survey}, for introductions to \emph{spectral graph theory}). Because this situation arises frequently, we briefly motivate this terminology and relate it to the analogous operators from the classical signal processing literature. 

For a function $f$ defined on the real line,
a \emph{Fourier multiplier operator} or \emph{filter} $\Psi$ reshapes the function's
frequencies through multiplication in the Fourier domain:
\begin{equation*}
\widehat{\Psi f}(\omega) = g(\omega)\hat{f}(\omega),\hbox{ for every frequency } \omega.
\end{equation*}
%Equivalently,
%denoting the Fourier and inverse Fourier transforms by ${\cal F}$ and ${\cal F}^{-1}$, we have
%we can write this as
Taking an inverse Fourier transform yields
\begin{align}\label{Eq:multiplier_def}
\Psi f(x) &= {\cal F}^{-1}\Bigl(g(\omega){\cal F}(f)(\omega)\Bigr)(x) \\
&= \frac{1}{2\pi}\int\limits_{\Rbb}g(\omega) \hat{f}(\omega) e^{i\omega x}~d\omega. \nonumber
\end{align}

Denoting the eigenvectors of $\L$ by $\left\{\boldsymbol{\chi}_{\l}\right\}_{\l=0,1,\ldots,N-1}$, we can extend this straightforwardly to functions defined on the vertices of a graph 
by replacing the Fourier transform and its inverse in \eqref{Eq:multiplier_def} with the graph Fourier transform 
%\begin{equation}\label{Eq:graph_FT}
$\hat{f}(\l) := \langle \mathbf{f},\boldsymbol{\chi_{\l}}\rangle = \sum_{n=1}^N  f(n)\chi^*_{\l}(n)$,
%\end{equation}
and its inverse
%\begin{equation}\label{Eq:graph_IFT}
$f(n) = \sum_{\l=0}^{N-1} \hat{f}(\l) \chi_{\l}(n).$
%\end{equation}
Namely, a %\emph{
graph Fourier multiplier operator is a linear operator $\Psi: \Rbb^N \rightarrow \Rbb^{N}$ that can be written as
\begin{align} \label{Eq:gfm_def}
\mathbf{\Psi f}(n) &= {\cal F}^{-1}\Bigl(g(\lambda_{\l}){\cal F}(f)(\l)\Bigr)(n) \nonumber \\
& = \sum\limits_{\l=0}^{N-1}g(\lambda_{\l}) \hat{f}(\l) \chi_{\l}(n).
\end{align}
We refer to $g(\cdot)$ as the \emph{multiplier} or \emph{graph spectral filter}.\footnote{Unlike \cite{shuman2013emerging}, we omit the hat symbol ($\hat{~}$) on the multiplier $g(\cdot)$, in order to maintain consistency with the notation most commonly used for matrix functions.} 
Equivalently, borrowing the above notation from the theory of matrix functions \cite{higham}, we can write 
\begin{align*}
\mathbf{\Psi}=g(\L) %: 
=\sum_{\l=0}^{N-1} g(\lambda_{\l})\boldsymbol{\chi}_{\l}\boldsymbol{\chi}_{\l}^*=\boldsymbol{\chi} g(\boldsymbol{\Lambda})\boldsymbol{\chi}^*. %,
\end{align*}
A high-level intuition behind %
graph spectral filtering \eqref{Eq:gfm_def}  is as follows. The eigenvectors corresponding to the lowest eigenvalues of the graph Laplacian are the ``smoothest'' in the sense that $\left|\chi_{\l}(m)-\chi_{\l}(n)\right|$ is small for neighboring vertices $m$ and $n$. 
%At the extreme is $\boldsymbol{\chi}_{0}$, which is a constant vector ($\chi_{0}(m)=\chi_{0}(n)$ for all $m$ and $n$). 
The inverse graph Fourier transform %\eqref{Eq:graph_IFT} 
provides a representation of a signal $\mathbf{f}$ as a superposition of the orthonormal set of eigenvectors of the graph Laplacian.
 %that form a basis for $\Rbb^N$.
The effect of
the graph Fourier multiplier operator $\mathbf{\Psi}$ is to modify
%modifies
the contribution of each eigenvector. %As an extreme
For example, applying a multiplier $g(\cdot)$ that is 1 for all $\lambda_{\l}$ below some threshold, and 0 for all $\lambda_{\l}$ above the threshold is equivalent to projecting the signal onto the eigenvectors of the graph Laplacian associated with the lowest %$k$
eigenvalues. %, for some $k$.
This is analogous to ideal lowpass filtering in the continuous domain. %Sections \ref{Se:applications}
Section %\ref{Se:smoothing}-\ref{Se:semi}
\ref{Se:applications} contains further intuition about and examples of graph Fourier multiplier operators. For more properties of the graph Laplacian eigenvectors, see \cite{shuman2013emerging} and \cite{lap_eigen}, and references therein.

\section{Illustrative Distributed Signal Processing Applications} \label{Se:applications}
In this section, we show that a number of distributed signal processing tasks can be solved as applications of graph multiplier operators or unions of graph multiplier operators.

\subsection{Denoising with Distributed Tikhonov Regularization} \label{Se:denoising_smooth}
First, we consider the distributed denoising task discussed in Section \ref{sec:intro}. %To recall, w
We start with a noisy signal $\mathbf{y}\in\R^N$ that is defined on a graph of $N$ sensors and has been corrupted by uncorrelated additive Gaussian noise. Through an iterative process of local communication and computation, %we would like 
each sensor should end up with a denoised estimate of its component, $f_n^0$, of the true underlying signal, $\mathbf{f}^0$.

To solve this problem, we enforce \emph{a priori} information that the target signal is smooth with respect to the underlying graph topology.
To enforce the global smoothness prior, we consider the class of regularization terms
$\mathbf{f}^{\transpose}\L^r \mathbf{f}$ for $r \geq 1$. The resulting distributed regularization problem has the form
\begin{eqnarray}\label{Eq:reg_prob}
\argmin_{\mathbf{f}} \frac{\tau}{2}\norm{\mathbf{f}-\mathbf{y}}_2^2+\mathbf{f}^{\transpose}\L^r \mathbf{f}.
\end{eqnarray}
Intuitively, the regularization term $\mathbf{f}^{\transpose}\L^r \mathbf{f}$ is small when the signal $\mathbf{f}$ has similar values at neighboring vertices with large weights (i.e., it is smooth). For example, when $r=1$, 
\begin{eqnarray*}
\mathbf{f}^{\transpose}\L \mathbf{f}=\frac{1}{2}\sum_{n \in \V}\sum_{m \sim n}\mathbf{W}_{m,n}\bigl(f_m-f_n\bigr)^2.  %, =\sum_{(u,v)\in {\cal E}}w(u,v)\left[f(v)-f(u)\right]^2
\end{eqnarray*}
%To see intuitively why incorporating such a regularization term into the objective function encourages smooth signals (with $r=1$ as an example), note that $\mathbf{f}^{\transpose}\L \mathbf{f}=0$ if and only if $\mathbf{f}$ is constant across all vertices, and, more generally,
%\begin{eqnarray*}
%\mathbf{f}^{\transpose}\L \mathbf{f}=\frac{1}{2}\sum_{n \in \V}\sum_{m \sim n}\mathbf{W}_{m,n}\bigl(f_m-f_n\bigr)^2, %=\sum_{(u,v)\in {\cal E}}w(u,v)\left[f(v)-f(u)\right]^2
%\end{eqnarray*}
%so $\mathbf{f}^{\transpose}\L \mathbf{f}$ is small when the signal $\mathbf{f}$ has similar values at neighboring vertices with large weights (i.e., it is smooth).

The proof of the following proposition
%Proposition \ref{Prop:reg} 
is included in the Appendix. 
\begin{proposition} \label{Prop:reg}
The solution to \eqref{Eq:reg_prob} is given by $\mathbf{Ry}$, where $\mathbf{R}$ is a graph Fourier multiplier operator of the form \eqref{Eq:gfm_def}, with multiplier $g(\lambda_{\l})=\frac{\tau}{\tau+2\lambda_{\l}^r}$ .\footnote{This filter $g(\lambda_{\l})$ is the graph analog of a first-order Bessel filter 
from classical signal processing of functions on the real line.}
\end{proposition}

So, one way to do distributed denoising is 
to approximately compute ${\mathbf{R}}\mathbf{y}$ in a distributed manner. We discuss methods to do this in Sections \ref{Se:chebyshev} and \ref{Se:other_methods}, and numerical examples are included in Section \ref{Se:basic_numerical} and Section \ref{Se:num_dist}. % \cite[Section V.B]{shuman_DCOSS_2011}.

\subsection{Distributed Smoothing} \label{Se:smoothing}
%Perhaps the simplest example application %of the distributed Chebyshev approximation method 
An application closely related to distributed denoising is distributed smoothing. Here, 
the graph Fourier multiplier is the \emph{heat kernel} $g(\lambda_{\l})=e^{-t\lambda_{\l}}$. In other words, a signal $\mathbf{y}\in \Rbb^N$ is smoothed by computing $\mathbf{H}_t y$, where 
%\begin{eqnarray*}
$(\mathbf{H}_t \mathbf{y})(n):=\sum_{\l=0}^{N-1} e^{-t\lambda_{\l}}\hat{y}(\l)\chi_{\l}(n)$ for fixed $t$.
%\end{eqnarray*}
%$\mathbf{H}_t$ clearly satisfies our definition of a graph Fourier multiplier operator. % (with $\eta=1$).
In the context of a centralized image smoothing application, \cite{hancock} discusses in detail the heat kernel %$\mathbf{H}_t$, 
and its relationship to classical Gaussian filtering. Similar to both the example at the end of Section \ref{Se:gfmo} and distributed Tikhonov regularization, the main idea is %once again
that the multiplier $g(\lambda_{\l})=e^{-t\lambda_{\l}}$ acts as a lowpass filter that attenuates the higher frequency (less smooth) components of %the signal
$\mathbf{y}$. The distributed smoothing problem is to compute $\mathbf{R}\mathbf{y}$, with $\mathbf{R}=\mathbf{H}_t=e^{-t \L}$ and each vertex $n$ beginning with only its observation $y_n$.

\subsection{Distributed Inverse Filtering} %Deconvolution}
Next, we consider the situation where node $n$ observes the $n^{\mathrm{th}}$ component of
$\mathbf{y}=\mathbf{\Psi f} + \boldsymbol{\nu}$, where $\mathbf{\Psi}$ is a graph Fourier multiplier operator with multiplier $g_{{\Psi}}(\cdot)$, and $\boldsymbol{\nu}$ is uncorrelated Gaussian noise. The task of the network is to recover $\mathbf{f}$ by inverting the effect of the %convolution
graph multiplier operator $\mathbf{\Psi}$. This is the distributed graph analog to the deblurring problem in imaging, which is discussed in \cite[Chapter 7]{peyre_book}. As discussed in \cite[Chapter 7]{peyre_book}, trying to recover $\mathbf{f}$ by simply applying the inverse filter in the graph Fourier domain, i.e., setting
\begin{align}\label{Eq:deconvolution_regu}
f_*(n)&=\sum_{\l=0}^{N-1}\left(\frac{1}{g_{{\Psi}}(\lambda_{\l})}\right)\hat{y}(\l)\chi_{\l}(n) \nonumber \\
&=\sum_{\l=0}^{N-1}\left(\hat{f}(\l)+\frac{\hat{\nu}(\l)}{g_{{\Psi}}(\lambda_{\l})}\right)\chi_{\l}(n) \nonumber \\
&=f(n)+\sum_{\l=0}^{N-1}\left(\frac{\hat{\nu}(\l)}{g_{{\Psi}}(\lambda_{\l})}\right)\chi_{\l}(n),
\end{align}
does not work well when $g_{{\Psi}}(\cdot)$ is zero (or close to zero) for high frequencies, because the summation in \eqref{Eq:deconvolution_regu} blows up, dominating $f(n)$.
Therefore, we again use the prior that the signal is smooth with respect to the underlying graph structure, and approximately solve the regularization problem
\begin{align}\label{Eq:deconvolution}
\argmin_{\mathbf{f}} \frac{\tau}{2}\norm{\mathbf{y}-\mathbf{\Psi f}}_2^2+\mathbf{f}^{\transpose}\L^r \mathbf{f}
\end{align}
in a distributed manner.
\begin{proposition} \label{Prop:deconvolution}
The solution to \eqref{Eq:deconvolution} is given by $\mathbf{Ry}$, where $\mathbf{R}$ is a graph Fourier multiplier operator with multiplier \begin{align*}
h(\lambda_{\l})&=\frac{\tau g_{{\Psi}}(\lambda_{\l})}{\tau g_{{\Psi}}^2(\lambda_{\l}) +2\lambda_{\l}^r}.
\end{align*}
\end{proposition}
The proof of Proposition \ref{Prop:deconvolution} is included in the Appendix.

\subsection{Distributed Semi-Supervised Classification}\label{Se:semi}
The goal of \emph{semi-supervised classification} is to learn a mapping from the data points $X=\{x_1,x_2,\ldots,x_N\}$ to
their corresponding labels $Y=\{y_1,y_2,\ldots,y_N\}$. The pairs $(x_i,y_i)$ are independently and identically sampled from a joint distribution $p(x,y)$
over the sample space ${\cal X} \times {\cal Y},$
where ${\cal Y}:=\{1,2,\ldots,\kappa\}$ is the space of $\kappa$ classes. The transductive classification problem is to use the full set of data points
$X=\{x_1,x_2,\ldots,x_N\}$ and the labels $Y_l=\{y_1,y_2,\ldots,y_l\}$ associated with a small portion of the data ($l\ll N$) to predict the labels
$Y_u=\{y_{l+1},y_{l+2},\ldots,y_N\}$ associated with the unlabeled data $X_u=\{x_{l+1},x_{l+2},\ldots,x_N\}$.

Many semi-supervised learning methods represent the data $X$ by an undirected, weighted graph, and then force the labels to be smooth with respect to the intrinsic structure of this graph.
We show how a number of these centralized graph-based semi-supervised classification methods can be 
written as applications of graph multiplier operators.  
Throughout, % the section, 
we assume there is one data point at each node in the graph, and the nodes know the weights of the edges connecting them to their neighbors in the graph. 
For example, each data point could be at a different node in a sensor network, and the weights
%with the graph weights***
could be a function of the physical distance between the nodes. 

For different choices of reproducing kernel Hilbert spaces (RKHS) ${\cal H}$, % on $\Rbb^N$,
a number of centralized semi-supervised classification methods estimate the label of the $n^{\mathrm{th}}$ data point ($n \in \{l+1,\ldots,N\}$) by
\begin{align}
\argmax_{j \in \{1,2,\ldots,\kappa\}} F_{nj}^{\mathrm{opt}},\hbox{ where~~~~~~~~~~~~} \label{Eq:semi_reg_1} \\
%\end{align}
%where $\mathbf{F}^{\mathrm{opt}}$ is the solution to
%\begin{align}\label{Eq:semi_reg}
%\hbox{where }
\mathbf{F}^{\mathrm{opt}}=\argmin_{\mathbf{F}\in \R^{N \times \kappa}} \sum_{j=1}^{\kappa} \left\{ \tau %\frac{\tau}{2}
\norm{\mathbf{F}_{:,j}-\mathbf{Y}_{:,j}}_2^2+\norm{\mathbf{F}_{:,j}}_{\cal H}^2 \right\}. \label{Eq:semi_reg}
\end{align}
In \eqref{Eq:semi_reg}, $\mathbf{A}_{:,j}$ denotes the $j^{\mathrm{th}}$ column of a matrix $\mathbf{A}$; $\mathbf{Y}$ is an ${N \times \kappa}$ matrix with entries
\begin{align*}
Y_{ij}=
\begin{cases}
1, &\mbox{if } i \in \{1,2,\ldots,l\} \mbox{ and the label for point }i\mbox{ is }j \\
0, &\mbox{otherwise}
\end{cases};
\end{align*}
and for some symmetric positive semi-definite matrix $\mathbf{S} \in \Rbb^{N \times N}$,
\begin{align}\label{Eq:H_inner}
\norm{\mathbf{f}}_{\cal H}^2=\langle \mathbf{f},\mathbf{f} \rangle_{{\cal H}}:=\langle \mathbf{f},\mathbf{S}\mathbf{f} \rangle=\mathbf{f}^{\transpose}\mathbf{S}\mathbf{f}.
\end{align}
Note that for any symmetric positive semi-definite matrix $\mathbf{S}$, ${\cal H}$ endowed with the inner product defined in \eqref{Eq:H_inner} is in fact a RKHS on $\mathbf{S}\Rbb^N$, and its kernel
is $k(i,j)=\left(\mathbf{S}^{-1}\right)_{ij}$, where $\mathbf{S}^{-1}$ denotes the pseudoinverse if $\mathbf{S}$ is not invertible \cite[Theorem 4]{smola}.

%We now give some examples of 
The following graph-based centralized semi-supervised classification methods %that 
fall into this category.
\begin{itemize}
\item In Tikhonov regularization, $\mathbf{S}=\L^r$ (e.g., \cite{belkin_matveeva})
\item Zhou \emph{et al.} \cite{zhou_bousquet} take $\mathbf{S}={\L^r_{\mathrm{norm}}}$, where $\L_{\mathrm{norm}}:=\mathbf{D}^{-\frac{1}{2}}\L\mathbf{D}^{-\frac{1}{2}}$ %, and also consider a variant with $\mathbf{P}=\L\mathbf{D}^{-1}$
%\item Zhou \emph{et al.} also consider a label propagation variant \cite{zhou_bousquet} where $\mathbf{P}=\L\mathbf{D}^{-1}$
\item Smola and Kondor~\cite{smola} consider a variety of kernel methods, including a diffusion process with $\mathbf{S}=\left[\exp\left(\frac{-\beta^2}{2} {\L_{\mathrm{norm}}} \right)\right]^{-1}$,
an inverse cosine with $\mathbf{S}=\left[\cos\left(\frac{\pi}{4} \L_{\mathrm{norm}}\right)\right]^{-1}$, and
an $r$-step random walk with $\mathbf{S}=\left(\beta \mathbf{I}_N-\L_{\mathrm{norm}}\right)^{-r}$, where $\beta\geq 2$ and
%and  where
%the $\sigma$'s are parameters and
$\mathbf{I}_N$ is the $N \times N$ identity matrix
%\footnote{All three of these $\mathbf{P}$
%matrices can be written as $\mathbf{P}=\sum_{\l=0}^{N-1}g(\lambda_{\l}) \boldsymbol{\chi}_l \boldsymbol{\chi}_l^{\transpose}$ for some $g(\cdot)$, where
%$\left\{\boldsymbol{\chi}_l\right\}_{\l=0,1,\ldots,N-1}$ are the eigenvectors of $\L_{\mathrm{norm}}$.}
\item Ando and Zhang's K-scaling method \cite{ando_zhang,johnson_zhang} takes
\begin{align*}
\mathbf{S}=(\gamma\mathbf{I}_N+\mathbf{D})^{-\frac{1}{2}}(\gamma\mathbf{I}_N+\L)(\gamma\mathbf{I}_N+\mathbf{D})^{-\frac{1}{2}},
\end{align*}
which reduces to $\L_{\mathrm{norm}}$ when $\gamma=0$.
\item Zhu \emph{et al.} \cite[Chapter 15]{chapelle} take the kernel approach a step further by solving a convex optimization problem to find a good $\mathbf{S}$
\end{itemize}

Before moving on to the distributed semi-supervised classification problem, we note that in all of the examples above, we can write
%can be written as 
$\mathbf{S}=h(\mathbf{P})$ for some $h(\cdot)$, where $\mathbf{P}$ is either the combinatorial graph Laplacian, the normalized graph Laplacian, or the matrix $\mathbf{S}$ used in the K-scaling method, all of which have the same sparsity pattern as $\L$ and are easily computable from the weighted adjacency matrix. 

Now, $\mathbf{F}^{\mathrm{opt}}$ in \eqref{Eq:semi_reg} can be equivalently rewritten as the solution to $\kappa$ separate minimization problems, with
\begin{align} \label{Eq:semi_reg_break}
\mathbf{F}_{:,j}^{\mathrm{opt}}&=\argmin_{\mathbf{f}\in \R^{N}} \left\{ \tau %\frac{\tau}{2}
\norm{\mathbf{f}-\mathbf{Y}_{:,j}}_2^2+\mathbf{f}^{\transpose}\mathbf{S}\mathbf{f} \right\} \nonumber \\
&=\argmin_{\mathbf{f}\in \R^{N}} \left\{ \tau %\frac{\tau}{2}
\norm{\mathbf{f}-\mathbf{Y}_{:,j}}_2^2+\mathbf{f}^{\transpose}h(\mathbf{P})\mathbf{f} \right\}.
\end{align}
We can write the solution to \eqref{Eq:semi_reg_break} as $\mathbf{R}\mathbf{Y}_{:,j}$, where $\mathbf{R}$ is a %generalized 
graph multiplier operator of the form outlined in Definition \ref{Def:gmo},
%\eqref{Eq:gen_graph_mult}, 
with respect to $\mathbf{P}$. The %corresponding 
optimal multiplier is $g(\lambda_{\l})=\frac{\tau}{\tau +h(\lambda_{\l})}$.

Therefore, the following is a method to distribute any of the centralized semi-supervised classification methods that can be written as \eqref{Eq:semi_reg_1} and \eqref{Eq:semi_reg}:
\begin{enumerate}
\item Node $n$ starts with or computes the entries of the $n^{\mathrm{th}}$ row of $\mathbf{P}$ %(which are easily computable from $\L$ in the cases mentioned above)
\item Each node $n$ forms the $n^{\mathrm{th}}$ row of $\mathbf{Y}$
\item For every $j\in\{1,2,\ldots,\kappa\}$, the nodes approximately compute $%\tilde
{\mathbf{F}}^{\mathrm{opt}}_{:,j}:=%\tilde
{\mathbf{R}}\mathbf{Y}_{:,j}$ in a distributed manner via algorithms outlined in the subsequent sections.
%Algorithm 1 (with $\mathbf{P}$ replacing $\L$)
\item Each node $n$ with an unlabeled data point computes its label estimate according to $\argmax_{j \in \{1,2,\ldots,\kappa\}} %\tilde
\left\{{F}_{nj}^{\mathrm{opt}}\right\}$
%\eqref{Eq:semi_reg_1}
\end{enumerate}
%By Propositions \ref{Prop:spec_bound} and \ref{Prop:smooth}, as we increase the Chebyshev approximation order $K$, the classification results of this distributed method converge to results of the corresponding centralized semi-supervised learning method.

\section{Distributed Chebyshev Polynomial Approximation of Graph Multiplier Operators} \label{Se:chebyshev}

Motivated by the fact that a number of distributed signal processing tasks can be viewed as applications of unions of graph multiplier operators, we proceed to the issue of how to approximately compute $\mathbf{\Phi f}$, where $\mathbf{\Phi}$ is of the form \eqref{Eq:union_def}, in a distributed setting. In this section, we introduce a computationally efficient approximation to unions of graph multiplier operators based on shifted Chebyshev polynomials. %

\subsection{The Centralized Chebyshev Polynomial Approximation}
Exactly computing $g(\mathbf{P})\mathbf{f}$
%$\mathbf{\Phi f}$
 requires explicit computation of the entire set of eigenvectors and eigenvalues of $\mathbf{P}$, which becomes computationally challenging as the size of the network, $N$, increases, even in a centralized setting. Druskin and Knizhnerman \cite{druskin} introduce a method to approximate $g(\mathbf{P})\mathbf{f}$ by $\tilde{g}(\mathbf{P})\mathbf{f}$, where $\tilde{g}(\cdot)$ is a polynomial approximation of $g(\cdot)$ computed by truncating a shifted Chebyshev series expansion of the function $g(\cdot)$ on the interval $[\lambda_{\min},\lambda_{\max}]$. Doing so circumvents the need to compute the full set of eigenvectors and eigenvalues of $\mathbf{P}$. This idea is extended to unions of graph Fourier multipliers in \cite[Section 6]{LTS-ARTICLE-2009-053}; that is, a computationally efficient approximation $\tilde{\mathbf{\Phi}}\mathbf{f}$ of $\mathbf{\Phi f}$ can be computed by approximating each %graph Fourier
multiplier $g_j(\cdot)$ by a truncated series of shifted Chebyshev polynomials.
We summarize this approach below.

For $y \in [-1,1]$, the Chebyshev polynomials $\left\{T_k(y)\right\}_{k=0,1,2,\ldots}$ are generated by
\begin{eqnarray*}
T_k(y):=\begin{cases}
1,&\hbox{ if }k=0 \\
y,&\hbox{ if }k=1 \\
2yT_{k-1}(y)-T_{k-2}(y),&\hbox{ if }k\geq 2
\end{cases}.
\end{eqnarray*}
These Chebyshev polynomials form an orthogonal basis for \\
$L^2\left([-1,1],\frac{dy}{\sqrt{1-y^2}} \right)$. So %that
every function $h$ on $[-1,1]$ that is square integrable with respect to the measure $dy/\sqrt{1-y^2}$ can be represented as
%\begin{eqnarray*}
$h(y)=\frac{1}{2}b_0+\sum_{k=1}^{\infty}b_k T_k(y)$,
%\end{eqnarray*}
where $\{b_k\}_{k=0,1,\ldots}$ is a sequence of Chebyshev coefficients that depends on $h(\cdot)$.
%for some sequence of Chebyshev coefficients $\{b_k\}_{k=0,1,\ldots}$ .
For a detailed overview of Chebyshev polynomials, including the above definitions and properties,
see \cite{handscomb}--\nocite{phillips}\cite{rivlin}.

By shifting the domain of the Chebyshev polynomials to $[0,\lambda_{\max}]$ via the transformation $x=\frac{\lambda_{\max}}{2}(y+1)$, we can represent each %graph Fourier
multiplier as
\begin{eqnarray} \label{Eq:multiplier_expansion}
g_j(x)=\frac{1}{2}c_{j,0}+\sum_{k=1}^{\infty}c_{j,k}\overline{T}_k(x), \hbox{ for all }x \in [0,\lambda_{\max}],
\end{eqnarray}
where $\overline{T}_k(x):=T_k\left(\frac{x-\alpha}{\alpha}\right)$, $\alpha:=\frac{\lambda_{\max}}{2}$, and 
\begin{align}\label{Eq:cheb_coeffs}
%&\overline{T}_k(x):=T_k\left(\frac{x-\alpha}{\alpha}\right),  \nonumber \\
%& \alpha:=\frac{\lambda_{\max}}{2}, \hbox{ and } \nonumber \\
& c_{j,k}:= \frac{2}{\pi}\int_{0}^{\pi}\cos(k\phi)~g_j\Bigl(\alpha \bigl(\cos(\phi) +1\bigr)\Bigr)~d\phi.
\end{align}
For $k \geq 2$, the shifted Chebyshev polynomials
satisfy
\begin{eqnarray*}
\overline{T}_k(x) = \frac{2}{\alpha}(x-\alpha)\overline{T}_{k-1}(x)
- \overline{T}_{k-2}(x).
\end{eqnarray*}
Thus, for any $\mathbf{f}\in \R^N$, we have
\begin{equation} \label{eq:cheby-f-recurrence}
  \overline{{T}}_k(\mathbf{P})\mathbf{f} =
  \frac{2}{\alpha}({\mathbf{P}}-{\alpha}\mathbf{I})\left( \overline{{T}}_{k-1}(\mathbf{P}) \mathbf{f}\right)
  - \overline{{T}}_{k-2}(\mathbf{P})\mathbf{f},
\end{equation}
where $\overline{{T}}_k(\mathbf{P}) \in \Rbb^{N \times N}$ and, by \eqref{Eq:operator_def}, the $n^{\mathrm{th}}$ element of $\overline{{T}}_k(\mathbf{P})\mathbf{f}$ is given by
\begin{eqnarray}\label{Eq:T_bar_def}
\left(\overline{{T}}_k({P}){f}\right)_n=\sum_{\l=0}^{N-1} \overline{T}_k(\lambda_{\l})\langle \mathbf{f},\mathbf{u}_{\l}\rangle
%\hat{f}(\l)
u_{\l}(n).
\end{eqnarray}

Now, to approximate the operator $\mathbf{\Phi}$, we can approximate each %graph Fourier
multiplier $g_j(\cdot)$ by the first $K+1$ terms in its Chebyshev polynomial expansion \eqref{Eq:multiplier_expansion}.
Then,
for every $j \in \{1,2,\ldots,\eta\}$ and $n \in \{1,2,\ldots,N\}$, we have
\begin{align} \label{Eq:cheby_approx}
&\left(\tilde{{\Phi}} {f}\right)_{(j-1)N+n} \nonumber \\
&\quad\quad := \left(\frac{1}{2}c_{j,0}{f}+\sum_{k=1}^{K}c_{j,k}\overline{{T}}_k({P}){f}\right)_n \\
&\quad\quad \stackrel{%(\ref{Eq:graph_IFT}), 
(\ref{Eq:T_bar_def})}= \sum_{\l=0}^{N-1}\left[\frac{1}{2}c_{j,0}+\sum_{k=1}^{K}c_{j,k}\overline{T}_k(\lambda_{\l})\right]\langle \mathbf{f},\mathbf{u}_{\l}\rangle
%\hat{f}(\l) 
u_{\l}(n) \nonumber \\
& \quad~\quad \approx \sum_{\l=0}^{N-1}\left[\frac{1}{2}c_{j,0}+\sum_{k=1}^{\infty}c_{j,k}\overline{T}_k(\lambda_{\l})\right]\langle \mathbf{f},\mathbf{u}_{\l}\rangle
%\hat{f}(\l) 
u_{\l}(n)  \nonumber \\
&\quad\quad \stackrel{(\ref{Eq:multiplier_expansion})} = \sum_{\l=0}^{N-1}g_j(\lambda_{\l}) 
\langle \mathbf{f},\mathbf{u}_{\l}\rangle
%\hat{f}(\l) 
u_{\l}(n) \nonumber
\\
&\quad~\quad \stackrel{(\ref{Eq:operator_def})} = \left({\Phi f}\right)_{(j-1)N+n} . \nonumber
\end{align}

To recap, we propose to compute $\tilde{\mathbf{\Phi}}\mathbf{f}$ by first computing the Chebyshev coefficients $\{c_{j,k}\}_{j=1,2,\ldots,\eta;~k=1,2,\ldots,K}$ according to \eqref{Eq:cheb_coeffs}, and then computing the sum in \eqref{Eq:cheby_approx}. The computational benefit of the Chebyshev polynomial approximation arises in \eqref{Eq:cheby_approx}
from the fact the vector $\overline{{T}}_k(\mathbf{P})\mathbf{f}$ can be computed recursively from $\overline{{T}}_{k-1}(\mathbf{P})\mathbf{f}$ and $\overline{{T}}_{k-2}(\mathbf{P})\mathbf{f}$ according to \eqref{eq:cheby-f-recurrence}.
The computational cost of doing so is dominated by the cost of %the 
matrix-vector multiplication with $\mathbf{P}$, 
%of the graph Laplacian $\L$, 
which is proportional to the number of edges, $|\E|$ \cite{LTS-ARTICLE-2009-053}. Therefore, if the underlying communication graph is sparse (i.e., $|\E|$ scales linearly with the network size $N$), it is far more computationally efficient to compute $\tilde{\mathbf{\Phi}}\mathbf{f}$ than $\mathbf{\Phi f}$. Finally, we note that in practice, setting the %Chebyshev 
approximation order $K$ to around 20 results in $\tilde{\mathbf{\Phi}}$ approximating $\mathbf{\Phi}$ 
closely enough for the applications we have examined.
%very closely in all of the applications we have examined.

\subsection{Distributed Computation of $\tilde{\mathbf{\Phi}}\mathbf{f}$} \label{Se:forward}
We now discuss the second benefit of the Chebyshev polynomial approximation: it is easily distributable. 
We consider the following scenario. There is a network of $N$ nodes, and each node $n$ begins with the following knowledge:
\begin{itemize}
\item $f(n)$, the $n^{\mathrm{th}}$ component of the signal $\mathbf{f}$
\item The identity of its neighbors, and the weights of the graph edges connecting itself to each of its neighbors
\item The Chebyshev coefficients, $c_{j,k}$, for $j \in \{1,2,\ldots,\eta\}$ and $k \in \{0,1,2,\ldots,K\}$. These can either be computed centrally according to \eqref{Eq:cheb_coeffs} and then transmitted throughout the network, or each node can begin with knowledge of the
multipliers, $\{g_j(\cdot)\}_{j=1,2,\ldots,\eta}$, and precompute the %first $M$
Chebyshev coefficients according to \eqref{Eq:cheb_coeffs}
\item An upper bound $\overline{\lambda_{\max}}$ on $\lambda_{\max}$, the largest eigenvalue of $\mathbf{P}$. %the graph Laplacian. 
This bound need not be %does not need to be
tight. For example, when $\mathbf{P}$ is the graph Laplacian $\L$, 
we can precompute a bound such as $\lambda_{\max}\leq \max\{d(m)+d(n); m \sim n \}$, where $d(n)$ is the degree of node $n$ \cite{anderson_morley}\cite[Corollary 3.2]{das}
\end{itemize}

The task is for each network node $n$ to compute
\begin{eqnarray}\label{Eq:ntarget}
\Big\{\Big(\tilde{{\Phi}}{f}\Big)_{(j-1)N+n}\Big\}_{j=1,2,\ldots,\eta}
\end{eqnarray}
by iteratively exchanging messages with its local neighbors in the network and performing some computations.

As a result of \eqref{Eq:cheby_approx}, for node $n$ to compute the desired sequence in \eqref{Eq:ntarget}, it suffices to learn $\left\{\left(\overline{{T}}_k({P}){f}\right)_n\right\}_{k=1,2,\ldots, K}$. %, where $M:=\max_j M_j$.
Note that $\left(\overline{{T}}_1({{P}}){f}\right)_n=\left(\frac{1}{\alpha}({{P}}-{\alpha}{I}){f}\right)_n$ and $P_{n,m}=0$ for all nodes $m$ that are not neighbors of node $n$. Thus, to compute $\left(\overline{{T}}_1({{P}}){f}\right)_n$, node $n$ just needs to receive $f(m)$ from all neighbors $m$. So once all nodes send their component of the signal to their neighbors, they are able to compute their respective components of $\overline{{T}}_1({\mathbf{P}})\mathbf{f}$. In the next step, each node $n$ sends the newly computed quantity $\left(\overline{{T}}_1({{P}}){f}\right)_n$ to all of its neighbors, enabling the distributed computation of $\overline{{T}}_2({\mathbf{P}})\mathbf{f}$ according to \eqref{eq:cheby-f-recurrence}. The iterative process of local communication and computation continues for $K$ rounds until each node $n$ has computed the required sequence $\left\{\left(\overline{{T}}_k({P}){f}\right)_n\right\}_{k=1,2,\ldots, K}$. In all,
$2K\card{{\E}}$ messages of length 1 are required for every node $n$ to compute its sequence of coefficients in \eqref{Eq:ntarget} in a distributed fashion. This distributed computation process is summarized in Algorithm 1. %\ref{alg1}.

\begin{algorithm}[t] \label{alg1}
\caption{Distributed Computation of $\tilde{\mathbf{\Phi}}\mathbf{f}$}
      Inputs at node $n$: $f_n$, $P_{n,m}~\forall m$, $\left\{c_{k,j}\right\}_{j=1,2,\ldots,\eta;~k=0,1,\ldots,K}$, \\
      and $\overline{\lambda_{\max}}$ \\
   Outputs at node $n$: $\left\{\left(\tilde{\Phi} f\right)_{(j-1)N+n}\right\}_{j=1,2,\ldots,\eta}$ \\ %[-2mm]
       \begin{algorithmic}[1]
       \STATE Set $\alpha = \frac{\overline{\lambda_{\max}}}{2}$
   \STATE Set $\left(\overline{{T}}_0({P}){f}\right)_n = f_n$
   \STATE Transmit $f_n$ to all neighbors ${\cal N}_n:=\{m:P_{n,m} \neq 0\}$
   \STATE Receive $f_m$ from all neighbors ${\cal N}_n$
   \STATE Compute and store
   \begin{align*}
   \left(\overline{{T}}_1({P}){f}\right)_n =
  \sum\limits_{m \in {\cal N}_n \cup n}\frac{1}{\alpha}P_{n,m}f_m  -f_n
  \end{align*}
   \FOR{$k=2,\ldots,K$}
   \STATE Transmit $\left(\overline{{T}}_{k-1}({P}){f}\right)_n$ to all neighbors ${\cal N}_n$
    \STATE Receive $\left(\overline{{T}}_{k-1}({P}){f}\right)_m$ from all neighbors ${\cal N}_n$
   \STATE Compute and store
   \begin{align*}
   \left(\overline{{T}}_k({P}){f}\right)_n =&
 \sum\limits_{m \in {\cal N}_n \cup n} \frac{2}{\alpha}P_{n,m}\left(\overline{{T}}_{k-1}({P}){f}\right)_m  \\
 &-2\left(\overline{{T}}_{k-1}({P}){f}\right)_n
  - \left(\overline{{T}}_{k-2}({P}){f}\right)_n
  \end{align*}
   \ENDFOR
   \FOR{$j \in \{1,2,\ldots,\eta\}$}
   \STATE Output
   \begin{align*}
   \left(\tilde{{\Phi}} {f}\right)_{(j-1)N+n} = \frac{1}{2}c_{j,0}f_n+\sum\limits_{k=1}^K c_{j,k} \left(\overline{{T}}_k({P}){f}\right)_n
   \end{align*}
   \ENDFOR
   \end{algorithmic}
   \end{algorithm}

An important point to emphasize again is that although the operator $\mathbf{\Phi}$ and its approximation $\tilde{\mathbf{\Phi}}$ are defined through the eigenvectors of $\mathbf{P}$, %the graph Laplacian, 
the Chebyshev polynomial approximation helps the nodes 
apply the operator to the signal without explicitly computing (individually or collectively) the eigenvalues or eigenvectors of $\mathbf{P}$, %the Laplacian, 
other than the upper bound on its spectrum. Rather, they initially communicate their component of the signal to their neighbors, and then communicate simple weighted combinations of the messages received in the previous stage in subsequent iterations. 
In this way, information about each component of the signal $\mathbf{f}$ diffuses through the network without direct communication between non-neighboring nodes.

\subsection{Distributed Computation of $\tilde{\mathbf{\Phi}}^*\mathbf{a}$ and $\tilde{\mathbf{\Phi}}^*\tilde{\mathbf{\Phi}}\mathbf{f}$} \label{Se:adj_d}
In some tasks, such as the distributed lasso presented in Section \ref{Se:lasso}, we not only need to apply unions of graph multiplier operators, but we also need to apply their adjoints.
The application of the adjoint $\tilde{\mathbf{\Phi}}^*$ of the Chebyshev polynomial approximate operator $\tilde{\mathbf{\Phi}}$ can also be computed in a distributed manner. Let
%\begin{eqnarray*}
$\mathbf{a}  = \left[ \mathbf{a}_1; \mathbf{a}_2; \ldots; \mathbf{a}_{\eta} \right] \in \Rbb^{\eta N}$,
%\end{eqnarray*}
where $\mathbf{a}_j \in \Rbb^N$. Then it is straightforward to show that
\begin{align}\label{Eq:adjoint_split}
\left(\tilde{{\Phi}}^*{a}\right)_n=\sum_{j=1}^{\eta}\left(\frac{1}{2}c_{j,0}{a}_j+\sum_{k=1}^{K}c_{j,k}\overline{{T}}_k(P){a}_j \right)_n.
\end{align}
We assume each node $n$ starts with knowledge of $a_j(n)$ for all $j \in \{1,2,\ldots,\eta\}$. For each $j \in \{1,2,\ldots,\eta\}$, the distributed computation of the corresponding term on the right-hand side of \eqref{Eq:adjoint_split} is done in an analogous manner to the distributed computation of $\tilde{\mathbf{\Phi}}\mathbf{f}$ discussed above. Since this has to be done for each $j$, $2K|\E|$ messages, each a vector of length $\eta$, are required for every node $n$ to compute $(\tilde{{\Phi}}^*{a})_n$. The distributed computation of $\tilde{\mathbf{\Phi}}^*\mathbf{a}$ is summarized in Algorithm 2.

\begin{algorithm}[t] \label{alg2}
\caption{Distributed Computation of $\tilde{\mathbf{\Phi}}^*\mathbf{a}$}
      Inputs at node $n$: $\left\{a_j(n)\right\}_{j=1,2,\ldots,\eta}$, $P_{n,m}~\forall m$, $\overline{\lambda_{\max}}$, \\ and $\left\{c_{k,j}\right\}_{j=1,2,\ldots,\eta;~k=0,1,\ldots,K}$, \\
   Output at node $n$: $\left(\tilde{{\Phi}}^*{a}\right)_n$\\ %[-2mm]
       \begin{algorithmic}[1]
       \STATE Set $\alpha=\frac{\overline{\lambda_{\max}}}{2}$
   \FOR{$j=1,2,\ldots,\eta$}
   \STATE Set $\left(\overline{{T}}_0(P){a}_j\right)_n = a_j(n)$
     \ENDFOR
   \STATE Transmit $\left\{a_j(n)\right\}_{j=1,2,\ldots,\eta}$ to all neighbors ${\cal N}_n:=\{m:P_{n,m} \neq 0\}$
   \STATE Receive $\left\{a_j(m)\right\}_{j=1,2,\ldots,\eta}$ from all neighbors ${\cal N}_n$
   \FOR{$j=1,2,\ldots,\eta$}
   \STATE Compute and store
    \begin{align*}
   \left(\overline{{T}}_1(\L){a}_j\right)_n =
  \sum\limits_{m \in {\cal N}_n \cup n}\frac{2}{\alpha}P_{n,m}a_j(m)  -2a_j(n)
  \end{align*}
\ENDFOR
   \FOR{$k=2,\ldots,K$}
    \STATE Transmit $\left\{\left(\overline{{T}}_{k-1}(P){a}_j\right)_n\right\}_{j=1,2,\ldots,\eta}$ to all neighbors ${\cal N}_n$
    \STATE Receive $\left\{\left(\overline{{T}}_{k-1}(P){a}_j\right)_m\right\}_{j=1,2,\ldots,\eta}$ from all neighbors ${\cal N}_n$
   \FOR{$j=1,2,\ldots,\eta$}
   \STATE Compute and store
   \begin{align*}
   \left(\overline{{T}}_k(P){a}_j\right)_n =&
 \sum\limits_{m \in {\cal N}_n \cup n} \frac{2}{\alpha}P_{n,m}\left(\overline{{T}}_{k-1}(P){a}_j\right)_m  \\
 &-2\left(\overline{{T}}_{k-1}(P)\mathbf{a}_j\right)_n
  - \left(\overline{{T}}_{k-2}(P)\mathbf{a}_j\right)_n
  \end{align*}
   \ENDFOR
   \ENDFOR
   \STATE Output
   \begin{align*}
    \left(\tilde{{\Phi}}^*{a}\right)_n=\sum_{j=1}^{\eta}\left\{\frac{1}{2}c_{j,0}a_j(n)+\sum_{k=1}^{K}c_{j,k}\left(\overline{{T}}_k(P){a}_j \right)_n\right\}.
   \end{align*}
   \end{algorithmic}
   \end{algorithm}

Using the property of the Chebyshev polynomials that
%\begin{eqnarray*}
$T_k(x)T_{k^{\prime}}(x)=\frac{1}{2}\left[T_{k+k^{\prime}}(x)+T_{|k-k^{\prime}|}(x) \right]$,
%\end{eqnarray*}
we can write
\begin{eqnarray*}
\left(\tilde{{\Phi}}^*\tilde{{\Phi}}{f}\right)_n=\left(\frac{1}{2}d_0 {f} + \sum_{k=1}^{2K} d_k \overline{{T}}_k(P){f} \right)_n.
\end{eqnarray*}
See \cite[Section 6.1]{LTS-ARTICLE-2009-053} for a similar calculation and an explicit formula for the coefficients $\left\{d_k\right\}_{k=0,1,\ldots,2K}$.
Thus, with each node $n$ starting with $f(n)$ as in Section \ref{Se:forward}, %the nodes can also compute 
$\tilde{\mathbf{\Phi}}^*\tilde{\mathbf{\Phi}}\mathbf{f}$ can be distributedly computed %in a distributed manner 
using $4K|\E|$ messages of length 1, with each node $n$ finishing with knowledge of $\left(\tilde{{\Phi}}^*\tilde{{\Phi}}{f}\right)_n$.

\subsection{Numerical Example} \label{Se:basic_numerical}
\begin{figure}[t]
\centering

\begin{minipage}[b]{0.45\linewidth}
   \centering
      \centerline{\small{Original}}
     \centerline{\includegraphics[width=1\linewidth]{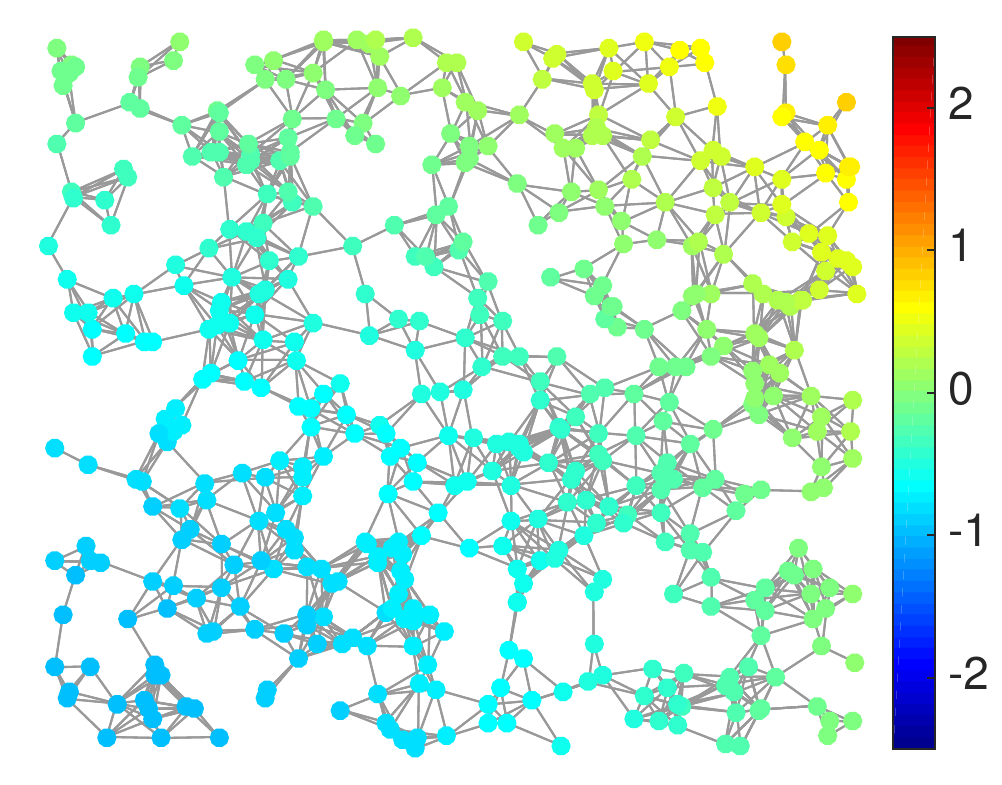}}
   \centerline{\small{(a)}}
\end{minipage}
\hfill
\begin{minipage}[b]{0.45\linewidth}
   \centering
         \centerline{\small{Noisy}}
      \centerline{\includegraphics[width=1\linewidth]{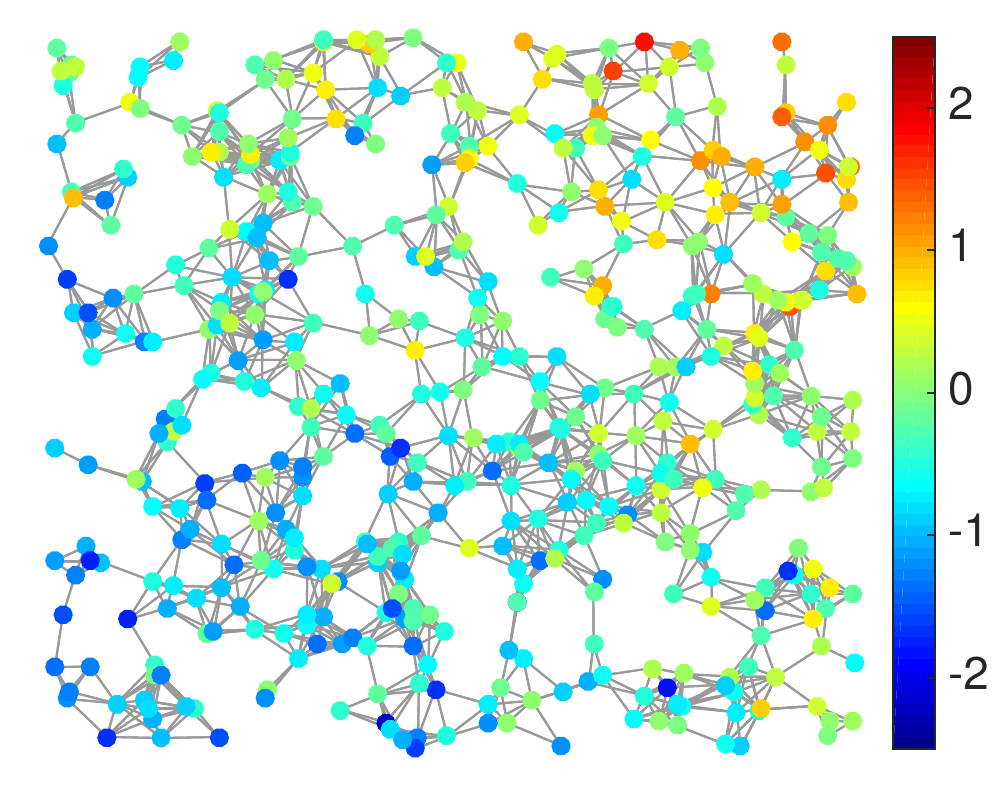}}
   \centerline{\small{(b)}}
   \end{minipage} \hspace{1.2in} \\
  % \hfill
  \vspace{0.4cm}
\begin{minipage}[b]{0.45\linewidth}
   \centering
         \centerline{\small{Denoised}}
         \vspace{0.1cm}
      \centerline{\includegraphics[width=1\linewidth]{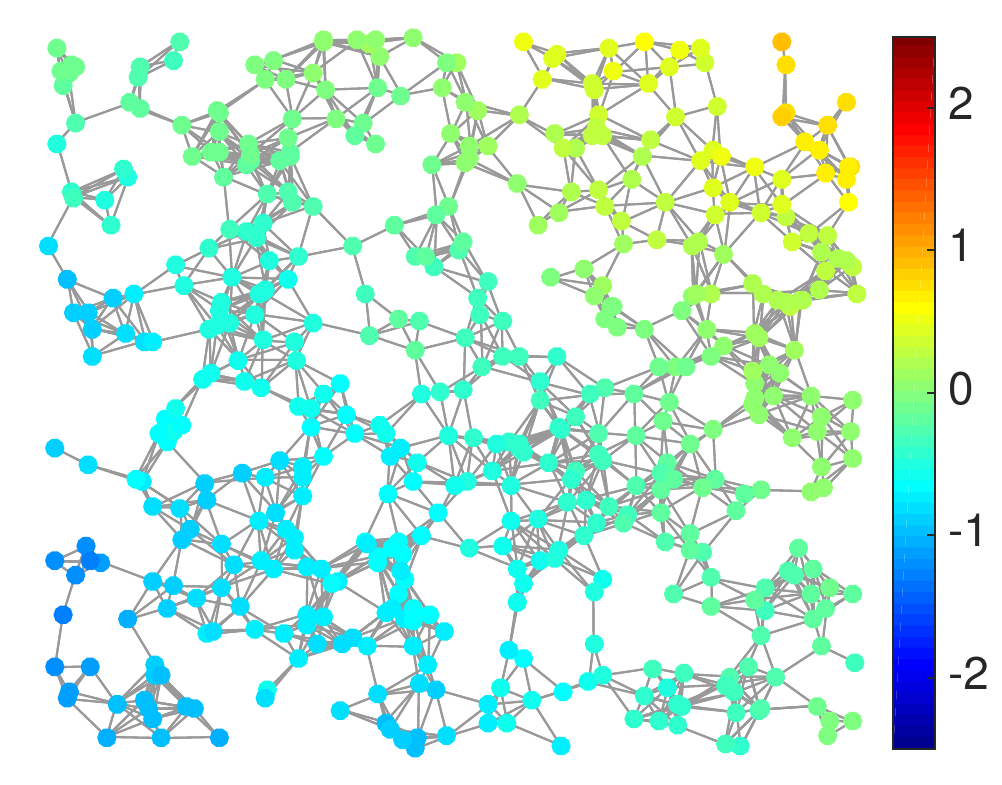}}
               \vspace{0.2cm}
   \centerline{\small{(c)}}
\end{minipage} %\\
%\vspace{0.4cm}
\hfill
%\hfill
\begin{minipage}[b]{0.48\linewidth}
   \centering
      \centerline{\includegraphics[width=1\linewidth]{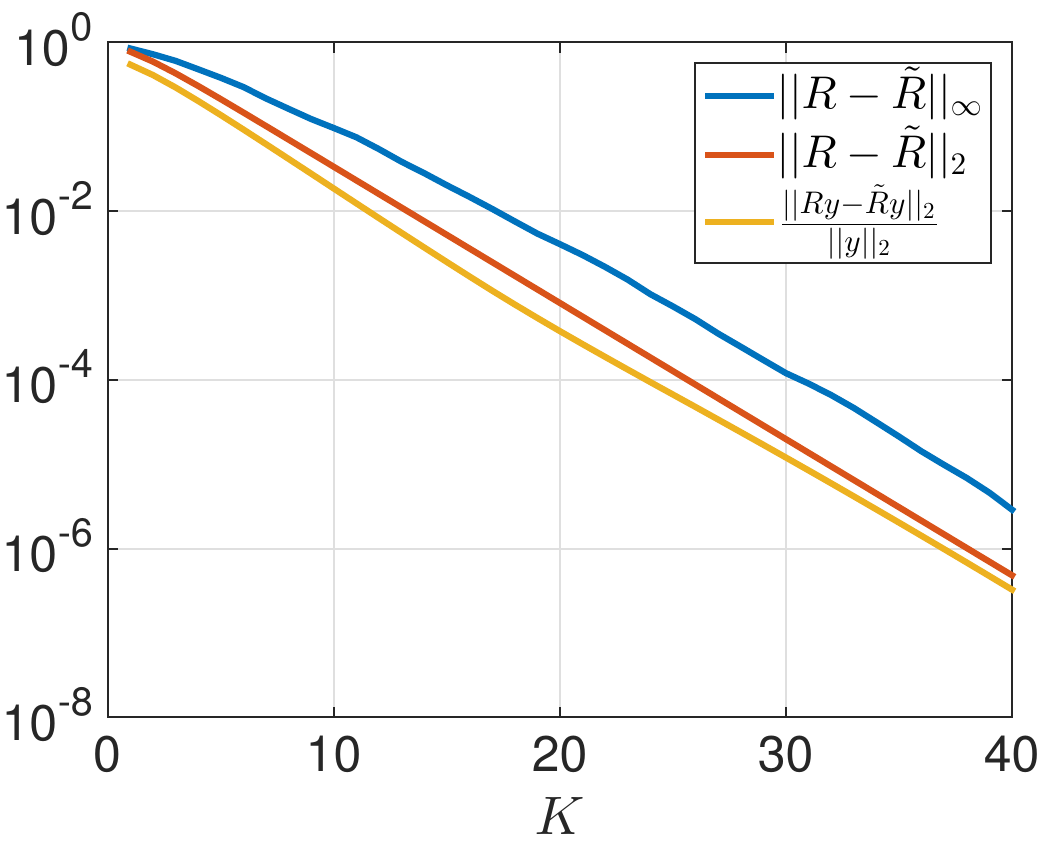}}
   \centerline{\small{~~(e)}}
\end{minipage} \hspace{1.2in} \\
\vspace{0.4cm}
\begin{minipage}[b]{.8\linewidth}
   \centering
   \centering
      \centerline{\includegraphics[width=1\linewidth]{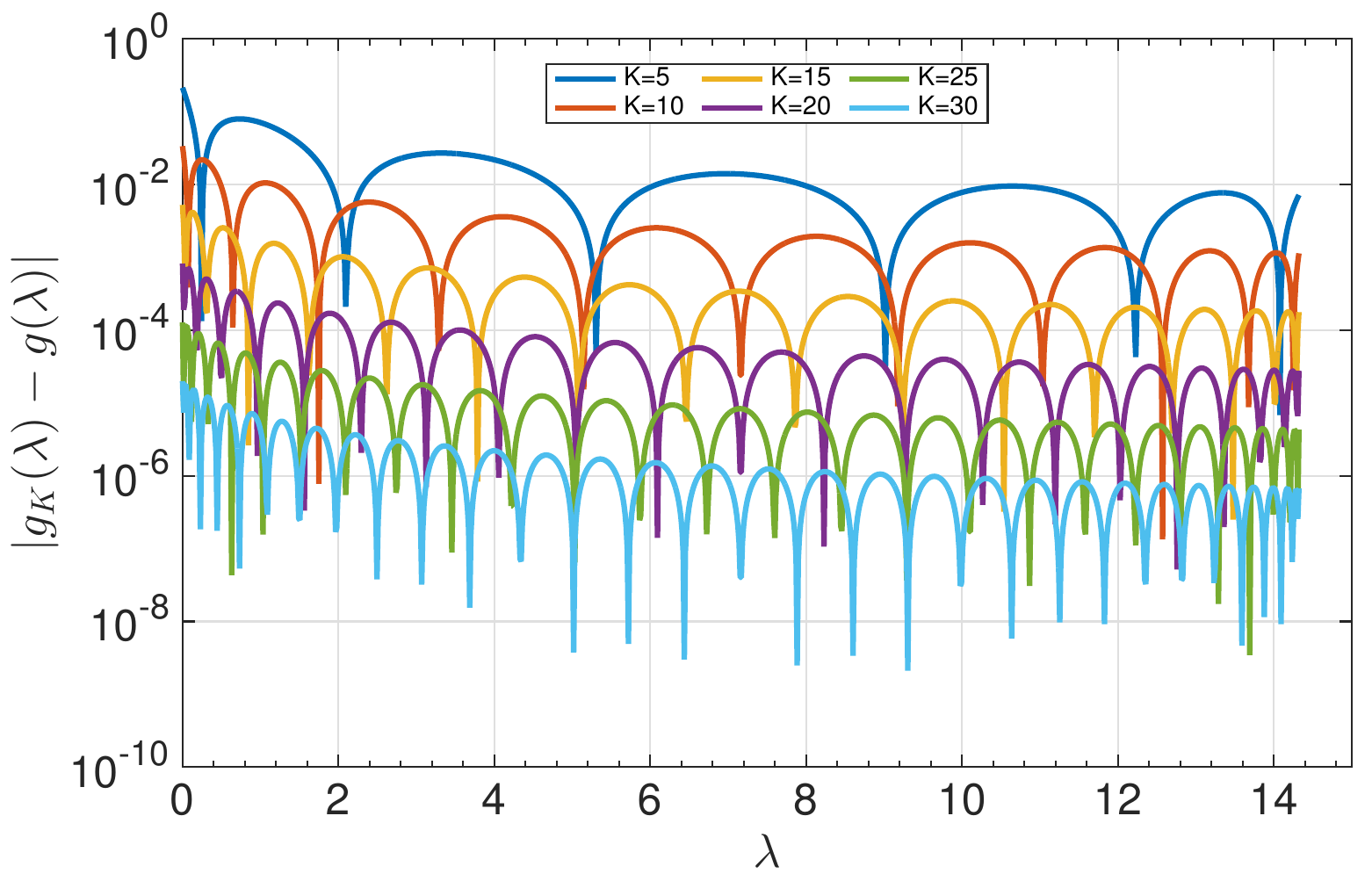}~~~~}
   \centerline{\small{~~~~(d)}} 
\end{minipage} 
\caption {Distributed denoising example. (a) The original signal with $h_n=n_x^2 +n_y^2-1$, where $n_x$ and $n_y$ are the $x$ and $y$ coordinates of sensor node $n$. (b) The noisy signal ${\bf y}$. (c) The denoised signal $\tilde{\bf R}{\bf y}$, the Chebyshev polynomial approximation (order $K=20$) to ${\bf R}{\bf y}=  \sum\limits_{\l=0}^{N-1}\frac{1}{1+2\lambda_{\l}} \hat{y}(\l) \chi_{\l}(n)$. (d) Approximation errors of shifted Chebyshev polynomial expansions of different orders for the filter $\hat{g}(\lambda_{\l})=\frac{1}{1+2\lambda_{\l}}$.  (e) Resulting approximation errors for the graph Fourier multiplier operator and graph filtered signal.} \label{Fig:network}
\end{figure}

We place 500 sensors randomly in the $[0,1] \times [0,1]$ square. 
We then construct a weighted graph according to %the 
a thresholded Gaussian kernel weighting function
based on the physical distance between nodes. %where t
The weight of edge $e$ connecting nodes $i$ and $j$ that are a distance $d(i,j)$ %from each other
apart is %given by
\begin{eqnarray*} %\label{Eq:gkw}
w(e)=
\begin{cases}
\exp\left({-\frac{[d(i,j)]^2}{2\theta^2}}\right) &\mbox{if } d(i,j) \leq \kappa \\
0 &\mbox{otherwise}
\end{cases},
\end{eqnarray*}
%for some parameters $\theta$ and $\kappa$.
with parameters $\sigma=0.074$ and $\kappa=0.075$. %, so that two sensor nodes are connected if %and only if
%their physical separation is less than %or equal to
%0.075. 
We create a smooth 500-dimensional signal with the $n^{\mathrm{th}}$ component given by $h_n = n_x^2 + n_y^2 -1$, where $n_x$ and $n_y$ are node $n$'s $x$ and $y$ coordinates in $[0,1] \times [0,1]$. Next, we corrupt each component of the signal ${\bf h}$ with uncorrelated additive Gaussian noise with mean zero and standard deviation 0.5, resulting in a noisy signal ${\bf y}$. Then we apply the graph Fourier multiplier operator $\tilde{\bf R}$, the Chebyshev polynomial approximation to ${\bf R}$ from Proposition \ref{Prop:reg}, with $\tau=r=1$ and $K=20$. The original signal ${\bf h}$, noisy signal ${\bf y}$, and denoised signal  $\tilde{\mathbf{R}}\mathbf{y}$ are shown in Figure \ref{Fig:network}(a)-(c). The Chebyshev polynomial approximation errors are shown in Figure  \ref{Fig:network}(d), and the resulting approximation errors for the graph Fourier multiplier operator and denoised signal are shown in Figure \ref{Fig:network}(e). We repeated this entire experiment 1000 times, with a new random graph and random noise each time, and the average mean square error for the denoised signals was 0.013, as compared to 0.250 average mean square error for the noisy signals.

\subsection{Approximation Error}
We %will 
use the following result, which bounds the spectral norm of the difference between a union of graph multiplier operators 
%(or union of such operators) 
and its Chebyshev polynomial approximation, %is used 
to analyze the distributed lasso problem in Section \ref{Se:lasso}. %\ref{Se:denoising_pw_smooth}. %\ref{Se:w_den_approx}.
\begin{proposition}\label{Prop:spec_bound}
Let $\mathbf{\Phi}$ be a union of $\eta$ %generalized 
graph multiplier operators; i.e., it has the form given in \eqref{Eq:union_def} for a real symmetric positive semi-definite matrix $\mathbf{P}$. 
Let $\tilde{\mathbf{\Phi}}$ be the order $K$ Chebyshev polynomial approximation of $\mathbf{\Phi}$. Define
\begin{align}\label{Eq:BK_def}
B(K):=\max_{j=1,2,\ldots,\eta} \left\{\sup_{\lambda \in [0,\lambda_{\max}]} \left\{\left|g_j(\lambda)-p_j^K(\lambda)\right| \right\}\right\},
\end{align}
where $\lambda_{\max}$ is the largest eigenvalue of $\mathbf{P}$, and $p_j^K(\cdot)$ is the order $K$ Chebyshev polynomial approximation of $g_j(\cdot)$. Then
\begin{align}\label{Eq:spec_bound}
\opnorm{{\mathbf{\Phi}}-\tilde{\mathbf{\Phi}}}_2:=\max_{\mathbf{f}\neq \mathbf{0}} \frac{\norm{({\mathbf{\Phi}}-\tilde{\mathbf{\Phi}})\mathbf{f}}_2}{\norm{\mathbf{f}}_2}\leq B(K)\sqrt{\eta}.
\end{align}
\end{proposition}

The proof of Proposition \ref{Prop:spec_bound} is included in the Appendix.

Finally, note that when the multipliers $g_j(\cdot)$ are smooth, the Chebyshev approximations $p_j^K(\cdot)$ converge to the multipliers rapidly as $K$ increases. The following proposition characterizes this convergence. %In particular, the following proposition applies.
\begin{proposition}[Theorem 5.14 in~\cite{handscomb}] \label{Prop:smooth}
If $g_j(\cdot)$ has $M+1$ continuous derivatives for all $j$, then $B(K)={\mathcal{O}}\left({K^{-M}}\right)$.
\end{proposition}

\section{Other Distributed Methods for Computing $g(\mathbf{P})\mathbf{y}$} \label{Se:other_methods}

In this section, we discuss some other methods for computing $g(\mathbf{P})\mathbf{y}$ in a distributed setting.
Most of these variations 
are not distributed computation methods \emph{per se}, but rather
centralized computational methods that can be 
distributed in the context of the applications mentioned above.

Higham~\cite[Chapter 13]{higham}, as well as Frommer and Simoncini \cite{frommer} provide excellent introductory overviews of centralized methods to compute $g(\mathbf{P})\mathbf{y}$ for large, sparse $\mathbf{P}$. Of the methods mentioned there, we do not consider contour integral or Krylov subspace methods,
which are not readily amenable to distributed computation. For example, in a distributed setting, the Lanczos method \cite{druskin,susnjara} would require a significant amount of extra communication at each iteration to compute vector norms. 

\subsection{Jacobi's Iterative Method}

For $\mathbf{S}=\mathbf{P}=\L_{\mathrm{\mathrm{norm}}}$, Zhou \emph{et al.} \cite{zhou_bousquet} propose to solve the semi-supervised classification problem \eqref{Eq:semi_reg} through the iteration
\begin{align}\label{Eq:iteration_Z}
\mathbf{F}^{(t+1)}=\frac{1}{1+\tau}\left[\left(\mathbf{I}_N-\mathbf{P}\right)\mathbf{F}^{(t)} + \tau \mathbf{Y} \right],~~~ \nonumber \\
~~~~~t=0,1,\ldots,T-1,
\end{align}
where $\mathbf{F}^{(0)}$ is arbitrary (set to $\mathbf{Y}$ in~\cite{zhou_bousquet}).\footnote{In \cite[Chapter 11]{chapelle}, similar iterative label propagation methods from \cite{zhu_g} and \cite{delalleau} are also
compared with the method of \cite{zhou_bousquet}.} The iteration \eqref{Eq:iteration_Z} is in fact just a particular instance of Jacobi's iterative method (see, e.g., \cite[Chapter 4]{saad}) to solve the set of linear equations
\begin{align}\label{Eq:lin_sys}
\left(\tau \mathbf{I}_N+\mathbf{P}\right)\mathbf{F}^{\mathrm{opt}}=\tau \mathbf{Y}.
\end{align}
So one alternative distributed semi-supervised classification method with $\mathbf{S}=\mathbf{P}=\L_{\mathrm{\mathrm{norm}}}$ is to compute the iterations \eqref{Eq:iteration_Z} in a distributed manner, with each node
starting with knowledge of its row of $\mathbf{P}$ and $\mathbf{Y}$. In fact, the communication cost of one iteration of \eqref{Eq:iteration_Z} is the same as the communication
cost of one iteration of the distributed computation of $\tilde{\mathbf{R}}\mathbf{Y}$ (lines 6 and 7 of Algorithm 1).

For graph multiplier operators whose multipliers have the property $g(\lambda_{\l})\neq 0$ for all $\l$,
%we can generalize 
the Jacobi method generalizes as follows. Suppose we wish to compute $\mathbf{Ry}$, where $\mathbf{R}$ is a graph multiplier operator with respect to $\mathbf{P}$ and with multiplier $g(\cdot)$. This is equivalent
to solving the linear system of equations
%\begin{align*}
$g(\mathbf{P})^{-1} \mathbf{x}=\mathbf{y}$.
%\Big(\sum_{\l=0}^{N-1}\frac{1}{g(\lambda_{\l})} \mathbf{u}_{l} \mathbf{u}^*_{l}\Big)\mathbf{f}=\mathbf{y}.
%\end{align*}
Assuming that the entries of the matrix $\mathbf{Q}=g(\mathbf{P})^{-1}$ are convenient to evaluate (e.g., for certain rational functions $g$), let
$\mathbf{Q}=\mathbf{Q}_D - \mathbf{Q}_O$, where $\mathbf{Q}_D$ contains the diagonal part of $\mathbf{Q}$. Then the Jacobi iteration is
\begin{align}\label{Eq:gen_gen_J}
\mathbf{x}^{(t+1)}=\mathbf{Q}_D^{-1}\mathbf{Q}_O \mathbf{x}^{(t)}+\mathbf{Q}_D^{-1}\mathbf{y},~t=0,1,\ldots,T-1.
\end{align}

One immediate drawback of Jacobi's method, as compared with the Chebyshev polynomial %our proposed 
method of Section \ref{Se:chebyshev}, is that it does not always converge. The iterations in
\eqref{Eq:gen_gen_J} converge for any $\mathbf{x}^{(0)}$ if and only if the spectral radius of $\mathbf{Q}_D^{-1}\mathbf{Q}_O$ is less than one \cite[Theorem 4.1]{saad}.
One sufficient condition for the latter to be true is that $\mathbf{Q}$ is strictly diagonally dominant, as is the case for example when $\mathbf{P}=\L$ and
$g(\lambda_{\l})=\frac{\tau}{\tau+\lambda_{\l}}$.
%${\color{red} 
Additionally, it may be too expensive computationally to evaluate the matrix $\mathbf{Q}$, or it may be a dense matrix, in which case the communication cost of a distributed method becomes prohibitive. For example, if $g=e^{-t\lambda}$, it is not efficient to fully evaluate $\mathbf{Q}$ and so this method 
is not applicable.

\subsection{Jacobi's Iterative Method with Chebyshev Acceleration} \label{Se:j_cheb}
When Jacobi's method does converge, we can accelerate \eqref{Eq:gen_gen_J} using the following algorithm \cite[Algorithm 6.7]{demmel}. Let $\rho$ be an upper bound on the spectral radius of
$\mathbf{Q}_D^{-1}\mathbf{Q}_O$, and define $\xi^{(0)}:=1$, $\xi^{(1)}:=\rho$, and $\mathbf{x}^{(1)}:=\mathbf{Q}_D^{-1}\mathbf{Q}_O \mathbf{x}^{(0)}+\mathbf{Q}_D^{-1}\mathbf{y}$. Then for $t \geq 1$, let
\begin{align} \label{Eq:J_Cheb_it}
\xi^{(t+1)}&=\frac{1}{\frac{2}{\rho \xi^{(t)}}-\frac{1}{\xi^{(t-1)}}},\hbox{ and } \nonumber \\
\mathbf{x}^{(t+1)} &=\frac{2\xi^{(t+1)}}{\rho \xi^{(t)}} \mathbf{Q}_D^{-1}\mathbf{Q}_O \mathbf{x}^{(t)} - \frac{\xi^{(t+1)}}{\xi^{(t-1)}} \mathbf{x}^{(t-1)}\nonumber \\
&\quad\quad + \frac{2 \xi^{(t+1)}}{\rho \xi^{(t)}} \mathbf{Q}_D^{-1} \mathbf{y}.
\end{align}
To distribute \eqref{Eq:J_Cheb_it}, each node $n$ must first learn $Q_{nn}$ and the $n^{\mathrm{th}}$ row of $\mathbf{Q}_O$.
For example, when $\mathbf{P}=\L_{\mathrm{\mathrm{norm}}}$ and $g(\lambda_{\l})=\frac{\tau}{\tau+\lambda_{\l}}$, as in \eqref{Eq:iteration_Z}, $Q_{nn}=\frac{\tau+1}{\tau}$ for all $n$, and the $n^{\mathrm{th}}$ row
of $\mathbf{Q}_O$ is just $-\frac{1}{\tau}$ times the $n^{\mathrm{th}}$ row of $\L_{\mathrm{\mathrm{norm}}}$.
%, which the nodes can easily compute in a distributed manner.
An additional challenge in a distributed setting may be to calculate the bound $\rho$.

%Finally, n
Note that while this method and %our 
the method of Section \ref{Se:chebyshev} share the same namesake, the use of the Chebyshev polynomials in the two is different. In Section \ref{Se:chebyshev}, we use Chebyshev
polynomials to approximate the multiplier, whereas this method improves the convergence speed of the Jacobi method by using Chebyshev polynomials to choose
the weights it uses to form the iterates in \eqref{Eq:J_Cheb_it} as
weighted linear combinations of the iterates in \eqref{Eq:gen_gen_J}. See Section 6.5.6 of \cite{demmel} for more details.

%%%%%%%%%%%%%%%%%%%%%%%%%%%%
% Moved up

\begin{figure*}[htb]
\centering
\begin{minipage}[b]{0.31\linewidth}
   \centering
   \centerline{\small{$\mathbf{S}=\L_{\mathrm{norm}}$}}
   \centerline{\includegraphics[width=1\linewidth]{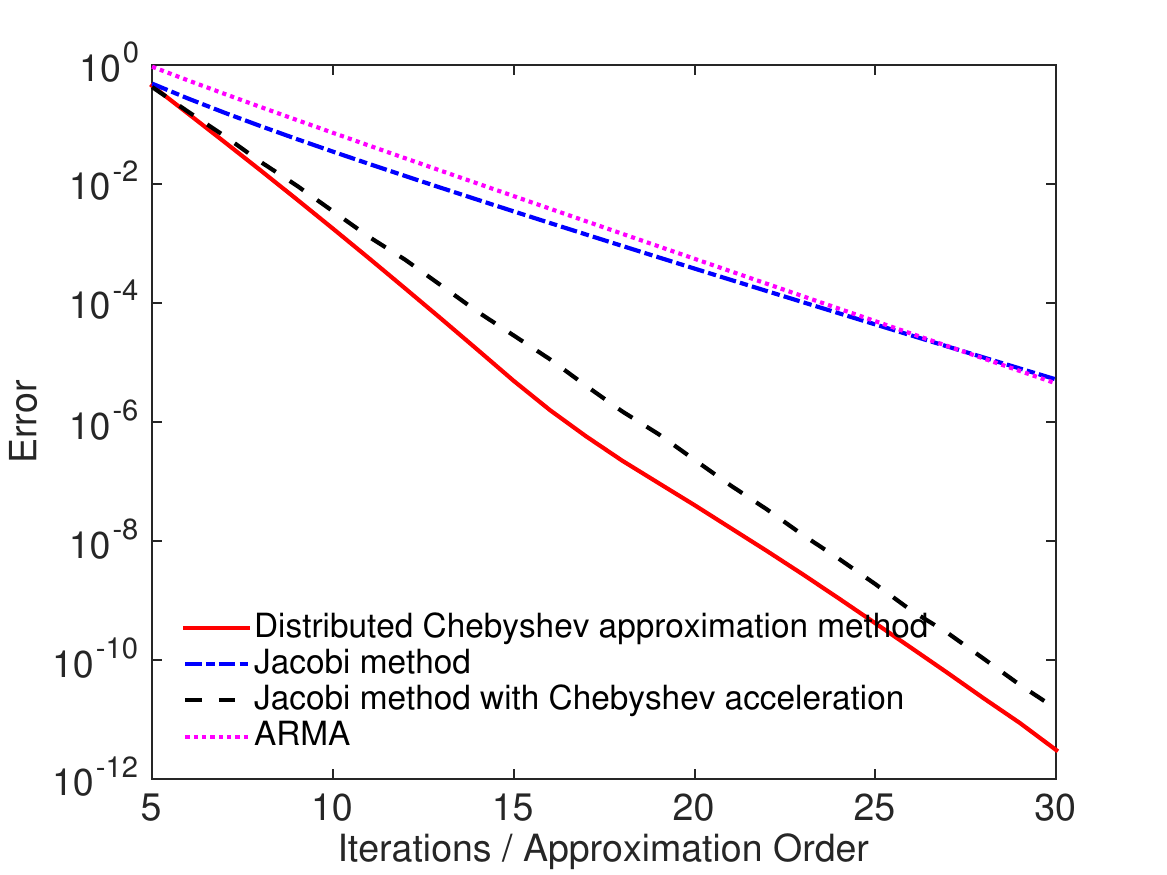}}
         \centerline{\small{(a)}}
\end{minipage}
\hfill
\begin{minipage}[b]{0.31\linewidth}
   \centering
   \centerline{\small{$\mathbf{S}=\L^2$}}
   \centerline{\includegraphics[width=1\linewidth]{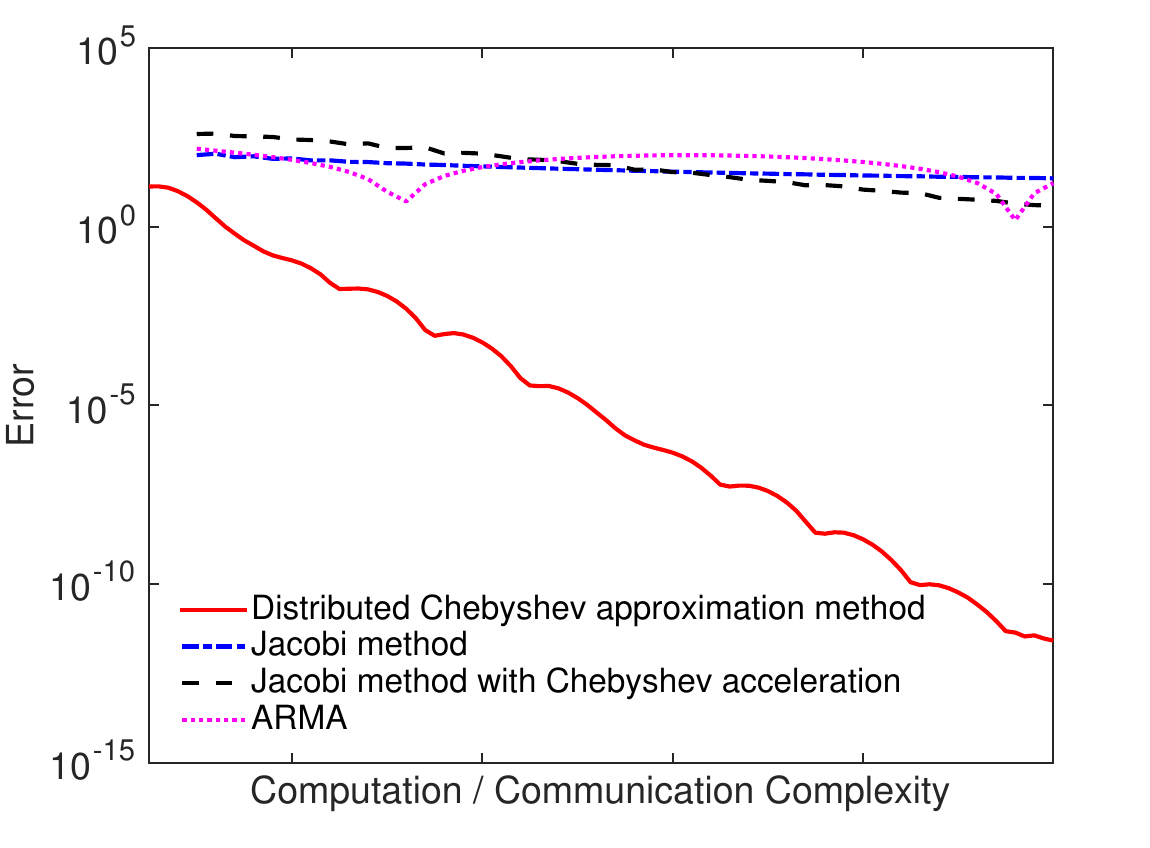}}
         \centerline{\small{(b)}}
\end{minipage} %\\
\hfill
\begin{minipage}[b]{0.31\linewidth}
   \centering
   \centerline{\small{$\mathbf{S}=\left(2\mathbf{I}_N-\L_{\mathrm{norm}}\right)^{-3}$}}
   \centerline{\includegraphics[width=1\linewidth]{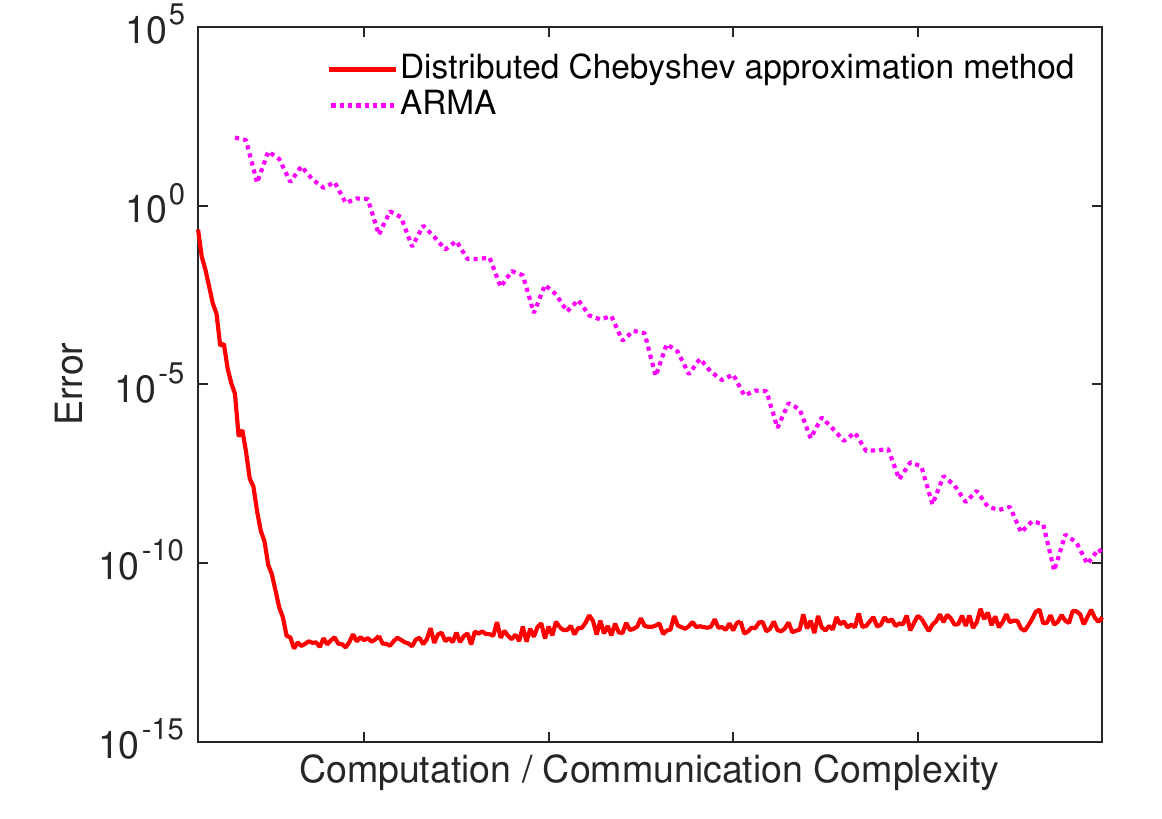}}
      \centerline{\small{(c)}}
\end{minipage}\\
\caption {Four different distributed methods to approximately compute $\mathbf{Ry}$, where $\mathbf{R}$ is a graph multiplier operator with respect to $\mathbf{P}$ for different choices of $\mathbf{P}$. In all cases,
the multiplier is $g(\lambda_{\l})=\frac{\tau}{\tau+h(\lambda_{\l})}$.  In (a), the error shown is $\norm{\mathbf{f}^{(K)}-\mathbf{f}}_2$, where $\mathbf{f}^{(K)}$  is either
$\tilde{\mathbf{R}}\mathbf{y}$ with an order $K$ approximation for our distributed Chebyshev approximation method, or the result of the $K^{\mathrm{th}}$ iteration for the Jacobi and ARMA methods. In (b) and (c), since iterations of the different algorithms have different computation and communication complexities, we normalize to a common scale.} \label{Fig:convergence}
\end{figure*}
%%%%%%%%%%%%%%%%%%%%%%%%%%%%%%

\subsection{Polynomial Approximation Variants}
Other orthogonal polynomials can also be used to generate %polynomial 
approximations via truncated expansions. For example, \cite{sheehan2010computing} uses Laguerre polynomials to approximate matrix exponentials. One advantage of this method is that Laguerre polynomials are orthonormal on $[0,\infty)$, so no upper bound on the spectrum is required. However, in the applications we consider, it is usually not hard to generate the upper bound $\overline{\lambda_{\max}}$.

%{\color{red}
In \cite{chen_saad}, Chen et al. first approximate the filter $g$ by a polynomial spline, and then compute orthogonal expansion coefficients of the spline in order to avoid the numerical integration involved in computing, e.g., the Chebyshev coefficients $\{c_k\}$ in \eqref{Eq:cheb_coeffs}. The conjugate residual-type algorithm of \cite{saad2006filtered} also uses the spline approach. However, in \cite{saad2006filtered}, the order $K+1$ polynomial approximation of a highpass filter $g$ takes the form  $\tilde{g}(\lambda)=\lambda \varphi(\lambda)$, where $\varphi$ is an order $K$ polynomial, forcing $\tilde{g}(0)$ to be equal to zero. If $g$ is a lowpass filter such as $g(\lambda)=e^{-\tau \lambda}$, then \cite{saad2006filtered} takes the approximation to be of the form $\tilde{g}(\lambda)=1-\lambda \varphi(\lambda)$, with $\varphi$ an order $K$ polynomial, once again guaranteeing zero approximation error at $\lambda=0$. This technique can be extended to bandpass filters by splitting the spectrum up into separate intervals, eventually leading to a three term recurrence with new weights that can be computed offline. A distributed implementation then carries the same communication cost as the single Chebyshev polynomial approximation.

\subsection{Rational Approximations} \label{Se:rational}
An alternative to a polynomial approximation is a rational approximation (see, e.g., \cite[Section 3.4]{frommer}) of the form 
\begin{align}\label{Eq:rational}
g(\lambda) \approx \frac{{\cal N}_\mu(\lambda)}{{\cal D}_\nu (\lambda)} =: \tilde{\tilde{g}}(\lambda),
\end{align}
where ${\cal N}_\mu$ and ${\cal D}_\nu$ are polynomials of degree $\mu$ and $\nu$, respectively. In the graph signal processing literature, references such as \cite{shi2015infinite,loukas2015distributed} refer to filters of the form \eqref{Eq:rational} as \emph{infinite impulse response filters}, since we can not write 
\begin{align}\label{Eq:FIR}
\tilde{\tilde{g}}(\mathbf{P}) = c_0 \mathbf{I} + \sum_{k=1}^K c_k \mathbf{P}^k,
\end{align}
for any choice of the order $K$ and series of coefficients $\{c_k\}$.

One benefit of rational approximations of the form \eqref{Eq:rational} is that they tend to provide better approximations than polynomials of lower orders, especially when $g$ features a singularity close to the spectrum of $\mathbf P$.  However, a major drawback is they tend to require extra subiterations, resulting in increased communication cost. For example, to compute $\mathbf{x}={\cal D}_\nu^{-1} (\mathbf{P})\mathbf{y}$, \cite{shi2015infinite} uses gradient descent to iteratively solve 
\begin{align} \label{Eq:gd}
 \argmin_{\mathbf{x}} ||{\cal D}_\nu (\mathbf{P}) \mathbf{x} - \mathbf{y}||^2.
 \end{align}
Yet,  \cite{shi2015infinite} estimates the number of iterations required to solve \eqref{Eq:gd} as $\frac{\max_{\l=0,1,\ldots,N-1}\{{\cal D}_\nu (\lambda_\l)^2\}}{\min_{\l=0,1,\ldots,N-1}\{{\cal D}_\nu (\lambda_\l)^2\}}$. Each of these iterations requires twice as much communication as the full distributed computation of an order $\nu$ matrix polynomial computation via Algorithm 1 (with $\eta=1$). So even when ${\cal N}_\mu$ and ${\cal D}_\nu$ are taken to be lower order polynomials, the communication requirements may still be significantly higher than a higher order polynomial approximation (where ${\cal D}_\nu (\lambda)=1$).

Some filters of the form \eqref{Eq:rational} with $\mu \leq K$ and $\nu =K$ can also be written as 
\begin{align}\label{Eq:parallel_arma}
\tilde{\tilde{g}}(\lambda)=\sum_{k=1}^K \frac{2r_k}{{\lambda_{\max}-\lambda_{\min}}-2\lambda-2p_k},
\end{align}
for some coefficient sequences $\{r_k\}$ and $\{p_k\}$. Loukas et al. \cite{loukas2015distributed} refer to such filters as parallel autoregressive moving average graph filters (ARMA) of order $K$, and show that if for all $k$, $|p_k|> \frac{\lambda_{\max}-\lambda_{\min}}{2}$, then  $\mathbf{x}=\tilde{\tilde{g}}(\mathbf{P})\mathbf{y}$ can be computed by iterating the following recursion for each term in the summation on the right-hand side of \eqref{Eq:parallel_arma}:
\begin{align}\label{Eq:arma_it}
\mathbf{x}_{k}^{(t+1)}&= \frac{1}{p_k}\left[\left(\frac{\lambda_{\max}-\lambda_{\min}}{2} \right)\mathbf{I} -\mathbf{P}\right]\mathbf{x}_{k}^{(t)} -\frac{r_k}{p_k} \mathbf{y}, \nonumber \\
&~~~~~~~~~~~~~~~~~~~~~~~~~~~~~~~~~t=0,1,\ldots,T-1
\end{align}
and then summing these results to find $\mathbf{x}=\sum_{k=1}^K \mathbf{x}_k$. Once again, these ARMA filters have the potential to yield a better approximation than a finite impulse response filter of the form on the right-hand side of \eqref{Eq:FIR} with the same order $K$; however, they require $T$ times the communication, where $T$ is the number of times one must iterate \eqref{Eq:arma_it} to convergence.

\subsection{Numerical Comparison}\label{Se:num_dist}

We consider the same random sensor network shown in Figure \ref{Fig:network}, and we generate
a signal $\mathbf{f}$ on the vertices of the graph with the components of $\mathbf{f}$ independently and identically sampled from a uniform distribution on $[-10,10]$. For different choices of $h(\cdot)$ and $\mathbf{P}=\L$ or $\mathbf{P}=\L_{\mathrm{norm}}$,
we define 
$$\mathbf{y}:=\left(\mathbf{I}_{500}+\frac{1}{\tau}h(\mathbf{P})\right) \mathbf{f}=g(\mathbf{P})\mathbf{f},$$ 
with $\tau=0.5$ and $g(\lambda)=\frac{\tau+h(\lambda)}{\tau}$.
Then, starting with $\mathbf{y}$, we iteratively compute an approximation to $\mathbf{f}$ in four different distributable ways: 1) $\tilde{\mathbf{R}}\mathbf{y}$, where
$\tilde{\mathbf{R}}$ is the Chebyshev approximation to $\mathbf{R}=g(\mathbf{P})^{-1}$;
2) with the Jacobi iteration \eqref{Eq:gen_gen_J}; 3) with the Jacobi iteration with Chebyshev acceleration \eqref{Eq:J_Cheb_it}; and 4) the ARMA iteration \eqref{Eq:arma_it}.

When $P=\L_{\mathrm{norm}}$ and $\mathbf{S}=h(\mathbf{P})=\L_{\mathrm{norm}}$,
%Since 
the filter $g^{-1}$ %in this example 
is the ratio of a constant and a first order polynomial, so we can take $K=1$ in \eqref{Eq:parallel_arma}. Taking the initial guess $\mathbf{x}^{(0)}$ to be $\mathbf{y}$ and $\lambda_{\min}=0$, the iteration \eqref{Eq:arma_it} becomes
\begin{align*}
\mathbf{x}^{(t+1)}&= \frac{2}{\tau+\lambda_{\max}}\left[\left(\frac{\lambda_{\max}}{2} \mathbf{I} -\mathbf{P}\right)\mathbf{x}^{(t)} +\tau\mathbf{y} \right]\\
&=\frac{2\tau}{\tau+\lambda_{\max}}\mathbf{y} + \frac{\lambda_{\max}}{\tau+\lambda_{\max}}\mathbf{x}^{(t)}-\frac{2}{\tau+\lambda_{\max}}\mathbf{P}\mathbf{x}^{(t)}.
\end{align*}
In this case, the communication requirements
of 
our method with Chebyshev approximation order $K$ are equal to the communication requirements of $T=K$ iterations of the latter three methods, so we plot the
errors $\norm{\mathbf{f}^{(K)}-\mathbf{f}}_2$ (where $\mathbf{f}^{(K)}$ corresponds to $\tilde{\mathbf{R}}\mathbf{y}$ with an order $K$ approximation in the first case or the result of the $K^{\mathrm{th}}$ iteration
in the latter three cases)  on the same axes in Figure \ref{Fig:convergence}(a).

When $P=\L$ and $\mathbf{S}=h(\mathbf{P})=\L^2$, computing $\mathbf{Q}_O \mathbf{x}^{(t)}$ in \eqref{Eq:gen_gen_J} and \eqref{Eq:J_Cheb_it} requires computing $\mathbf{W}\mathbf{x}^{(t)}$, which requires twice the communication and computation of a single iteration of Algorithm 1 for the polynomial approximation. For the ARMA approach, we can write the filter $g(\lambda)=\frac{\tau}{\tau+\lambda^2}$ exactly in the form of \eqref{Eq:parallel_arma} with 
$p_1=\sqrt{\tau}i+\frac{\lambda_{\max}}{2},~p_2=-\sqrt{\tau}i+\frac{\lambda_{\max}}{2},~r_1=-\frac{\sqrt{\tau}i}{2}$, and $r_2=\frac{\sqrt{\tau}i}{2}$.

When $P=\L_{\mathrm{norm}}$ and $\mathbf{S}=h(\mathbf{P})=\left(2\mathbf{I}_{500}-\L_{\mathrm{norm}}\right)^{-3}$ (a three-step random walk process), the Jacobi method does not converge. We have $h(\lambda)=(2-\lambda)^{-3}$, and thus
$$g(\lambda)=\frac{\tau}{\tau+h(\lambda_{\l})} = 1 - \frac{2}{(2-\lambda)^3+2},$$
the last term of which can be written as a third order ARMA filter.

Figure \ref{Fig:convergence} compares the approximation error to the communication/computation complexity for each of these methods and choices of $\mathbf{S}$. In these experiments, not only does our proposed method always converge, but it converges faster and with less communication than the alternative methods we tested.

\section{Distributed Lasso} \label{Se:lasso}

In Section \ref{Se:applications}, we presented a number of distributed signal processing tasks that could be represented as a single application of a union of graph multiplier operators. In this section, we present a distributed wavelet denoising example that requires repeated applications of unions of graph multiplier operators and their adjoints. Recall that the distributed Tikhonov regularization method from Section \ref{Se:denoising_smooth} is an efficient way to denoise a signal when we have \emph{a priori} information that the underlying signal is globally smooth. The distributed wavelet denoising method %here 
is better suited to situations where
we start with a prior belief that the signal is not globally smooth, but rather piecewise smooth, which corresponds to the signal being sparse in the spectral graph wavelet domain \cite{LTS-ARTICLE-2009-053}.

The spectral graph wavelet transform, %$\mathbf{\Phi}$, 
defined in \cite{LTS-ARTICLE-2009-053} is precisely of the form of $\mathbf{\Phi}$ in \eqref{Eq:operator_def}. Namely, it is composed of one %graph Fourier
multiplier, $h(\cdot)$, that acts as a lowpass filter to stably represent the signal's low frequency content, and $J$ wavelet operators, defined by $g_j(\lambda_{\l})=g(t_j \lambda_{\l})$, where $\{t_j\}_{j=1,2,\ldots,J}$ is a set of scales and $g(\cdot)$ is the wavelet multiplier %kernel
that acts as a bandpass filter.

The most common way to incorporate a sparse prior in a centralized setting is to regularize via a weighted version of the %\emph{lasso} \cite{lasso},
\emph{least absolute shrinkage and selection operator (lasso)} \cite{lasso},
also called \emph{basis pursuit denoising} \cite{basispursuit}:
\begin{align} \label{Eq:lasso}
%\hat{a}=\argmin_b~\frac{1}{2}\norm{y-W^*b}_2^2+\mu\norm{b}_1~,
\argmin_{\mathbf{a}}~\frac{1}{2}\norm{\mathbf{y}-\mathbf{\Phi}^*{\mathbf{a}}}_2^2+\norm{{\mathbf{a}}}_{1,\boldsymbol{\mu}}~,
\end{align}
where
%\begin{eqnarray*}
$\norm{{\mathbf{a}}}_{1,\boldsymbol{\mu}}:=\sum_{i=1}^{N(J+1)} \mu_i \left|{a}_i\right|$ and $\mu_i>0$ for all $i$.
%\end{eqnarray*}
%which
The optimization problem in \eqref{Eq:lasso} can be solved for example by iterative soft thresholding \cite{DDD}. The initial estimate of the wavelet coefficients ${\mathbf{a}}^{(0)}$ is arbitrary, and at each iteration of the soft thresholding algorithm, the update of the estimated wavelet coefficients is given by
\begin{eqnarray}\label{Eq:ista_update}
{a}_i^{(\beta)}=\Scal_{{\mu}_i {\gamma}}\left(\Bigl({{a}}^{(\beta-1)}+{\gamma}{\Phi}\left[{y}-{\Phi}^*{{a}}^{(\beta-1)}\right]\Bigr)_i\right),  \nonumber \\
i=1,2,\ldots,N(J+1);~\beta=1,2,\ldots
\end{eqnarray}
where ${\gamma}$ is the step size and $\Scal_{{\mu}_i{\gamma}}$ is the shrinkage or soft thresholding operator
\begin{eqnarray*}
\Scal_{{\mu}_i{\gamma}}(z):=\left\{
\begin{array}{ll}
0&,\mbox{ if } \mid z \mid \leq {\mu}_i{\gamma} \\
z-\mbox{sgn}(z){\mu}_i{\gamma}&, \mbox{ o.w.}
\end{array}
\right. .
\end{eqnarray*}
The iterative soft thresholding algorithm converges to ${{\mathbf{a}}}_*$, the minimizer of \eqref{Eq:lasso}, if ${\gamma} < \frac{2}{{\norm{\mathbf{\Phi}^*}^2}}$ \cite{combettes}. The final denoised estimate of the signal is then given by $\mathbf{\Phi}^*{{\mathbf{a}}}_*$.

We now turn to the issue of how to implement the above algorithm in a distributed fashion by sending messages between neighbors in the network. %he sensor nodes.
One option would be to use the distributed lasso algorithm of \cite{dlasso_conf,dlasso}, which is a special case of the alternating direction method of multipliers \cite[p.~253]{par_dist_opt_book}. In every iteration of that algorithm, each node transmits its current estimate of \emph{all} the wavelet coefficients to its local neighbors. %, and subsequently updates these estimates and some auxiliary variables. Thus, w
%With a transform the size of
With the spectral graph wavelet transform, %there are
that method requires $2\card{{\E}}$ total messages %exchanged
at every iteration, with each message being a vector of length  $N(J+1)$.
A method where the amount of communicated information does not grow with $N$ (beyond the number of edges, $\card{{\E}}$) would be highly preferable.

The %fast approximate
Chebyshev polynomial approximation of the spectral graph wavelet transform
%based on the Chebyshev polynomials
allows us to accomplish this goal.
Our approach, which is summarized in Algorithm 3, is to approximate $\mathbf{\Phi}$ by $\tilde{\mathbf{\Phi}}$, and use the distributed implementation of the approximate wavelet transform and its adjoint to
perform iterative soft thresholding in order to solve
\begin{align} \label{Eq:approx_lasso}
%\hat{a}=\argmin_b~\frac{1}{2}\norm{y-W^*b}_2^2+\mu\norm{b}_1~,
\argmin_{\tilde{\mathbf{a}}}~\frac{1}{2}\norm{\mathbf{y}-\tilde{\mathbf{\Phi}}^*\tilde{\mathbf{a}}}_2^2+\norm{\tilde{\mathbf{a}}}_{1,\boldsymbol{\mu}}.
\end{align}
In the first soft thresholding iteration, each node $n$ must learn $(\tilde{{\Phi}}{y})_{(j-1)N+n}$ at all scales $j$, via Algorithm 1.
These coefficients are then stored for future iterations. In the $\beta^{\mathrm{th}}$ iteration, each node $n$ must learn the $J+1$ coefficients of
$\tilde{\mathbf{\Phi}}\tilde{\mathbf{\Phi}}^*\tilde{\mathbf{a}}^{(\beta-1)}$ centered at $n$, by sequentially applying the operators
$\tilde{\mathbf{\Phi}}^*$ and $\tilde{\mathbf{\Phi}}$ in a distributed manner via Algorithms 2 and 1, %the methods of Sections \ref{Se:adj_d} and \ref{Se:forward},
respectively. %Finally, w
When a stopping criterion for the soft thresholding is satisfied,
%the soft thresholding iterations result in the satisfaction of a stopping criterion,
the adjoint operator $\tilde{\mathbf{\Phi}}^*$ is applied again in a distributed manner to the resulting coefficients $\tilde{{{\mathbf{a}}}}_*$, %via the method of Section \ref{Se:adj_d},
and node $n$'s denoised estimate of its signal is $\left(\tilde{{\Phi}}^*\tilde{{{{a}}}}_*\right)_n$. The stopping criterion may simply be a fixed number of iterations, or it may be
when $\left|\left(\tilde{{\Phi}}^*\tilde{a}^{(\beta)}\right)_n-\left(\tilde{{\Phi}}^*\tilde{a}^{(\beta-1)}\right)_n\right|<\epsilon$ for all $n$ and some small $\epsilon$.
Finally, note that we could also optimize the weights $\boldsymbol{\mu}$ by performing distributed cross-validation, as discussed in \cite{dlasso_conf,dlasso}.

We now examine the communication requirements of this approach. Recall from Section \ref{Se:forward} that
$2K\card{{\E}}$ messages of length 1 are required to compute $\tilde{\mathbf{\Phi}}\mathbf{y}$ in a distributed fashion.
Distributed computation %via \eqref{Eq:joint_op}
of $\tilde{\mathbf{\Phi}}\tilde{\mathbf{\Phi}}^*\tilde{\mathbf{a}}^{(\beta-1)}$, the other term needed in the iterative thresholding update \eqref{Eq:ista_update}, requires $2K\card{{\E}}$ messages of length $J+1$ and $2K\card{{\E}}$ messages of length $1$. % each a vector of length $J+1$.
 The final application of the adjoint operator $\tilde{\mathbf{\Phi}}^*$ to recover the denoised signal estimates requires another $2K\card{{\E}}$ messages, each a vector of length $J+1$. Therefore, the Chebyshev polynomial approximation to the spectral graph wavelet transform enables us to iteratively solve the weighted lasso %\eqref{Eq:lasso_app}
in a distributed manner where the communication workload only scales with the  size of the network through $\card{\E}$, and is otherwise independent of the network dimension $N$.

The reconstructed signal in Algorithm 3 is $\tilde{\mathbf{\Phi}}^*\tilde{\mathbf{a}}_*$, where $\tilde{\mathbf{a}}_*$ is the solution to the lasso problem \eqref{Eq:approx_lasso}. %, which features the approximate spectral graph wavelet transform $\tilde{\mathbf{W}}^*$. \
A natural question is how good of an approximation $\tilde{\mathbf{\Phi}}^*\tilde{\mathbf{a}}_*$ is to ${\mathbf{\Phi}}^*{\mathbf{a}}_*$, where ${\mathbf{a}}_*$ is the solution to the original lasso problem \eqref{Eq:lasso}. The following proposition bounds the squared distance between these two quantities by a term proportional to the spectral norm of the difference between the exact and approximate spectral graph wavelet operators.
\begin{proposition} \label{Prop:denoising}
$\norm{\tilde{\mathbf{\Phi}}^*\tilde{\mathbf{a}}_*-{\mathbf{\Phi}}^*{\mathbf{a}}_*}_2^2 \leq C \opnorm{\tilde{\mathbf{\Phi}}-\mathbf{\Phi}}_2$, where $\opnorm{\cdot}_2$ is the spectral norm, and the constant $C=\frac{\norm{\mathbf{y}}_2^3}{\min_i \mu_i}$. %depends on $\boldsymbol{\mu}$ and $\norm{\mathbf{y}}_2$.
\end{proposition}

Combining Proposition \ref{Prop:denoising}, whose proof is included in the Appendix, with \eqref{Eq:spec_bound}, we have
\begin{align}\label{Eq:denoising_bound}
\norm{\tilde{\mathbf{\Phi}}^*\tilde{\mathbf{a}}_*-{\mathbf{\Phi}}^*{\mathbf{a}}_*}_2^2 \leq \frac{\norm{\mathbf{y}}_2^3}{\min_i \mu_i} B(K) \sqrt{J+1}.
\end{align}
Thus, as we increase the approximation order $K$, $B(K)$ and the right-hand side of \eqref{Eq:denoising_bound} tend toward zero (at a speed dependent on the smoothness of the graph wavelet multipliers $g(\cdot)$ and %scaling function
$h(\cdot)$).

Finally, to illustrate the distributed lasso, we %again 
consider a numerical example. % where
We use the same 500 node sensor network as in Section \ref{Se:basic_numerical}.
This time, however, the underlying signal is %only 
piecewise smooth, but not globally smooth, with the $n^{\mathrm{th}}$ component given by
\begin{align*}
f_n^0=
\begin{cases}
-2n_x+0.5,& \hbox{ if } n_y \geq 1-n_x \\
n_x^2+n_y^2+0.5,& \hbox{ if } n_y < 1-n_x
\end{cases}.
\end{align*}
We %again 
corrupt each component of the signal $\mathbf{f}^0$ with uncorrelated additive Gaussian noise with mean zero and standard deviation 0.5. We then solve problem \eqref{Eq:approx_lasso} in a distributed manner using Algorithm 3. %In \eqref{Eq:approx_lasso}, w
We use a spectral graph wavelet transform with 6 wavelet scales, implemented by the Graph Signal Processing Toolbox \cite{gspbox}.
In Algorithm 3, we run 300 soft thresholding iterations and take $\gamma=0.2$,  $\mu_i=0.75$ for all the wavelet coefficients, and $\mu_i=0.01$ for all the scaling coefficients.\footnote{The scaling coefficients in the spectral graph wavelet transform are not expected to be sparse.}
We do not perform any distributed cross-validation to optimize the weights $\boldsymbol{\mu}$. We repeated this entire experiment 1000 times, with
a new random graph and random noise each time.\footnote{The reported errors are averaged over the 441 random graph realizations that were connected.} The
average mean square errors were 0.250 for the noisy signals, 0.098 for the estimates produced by the Tikhonov regularization method \eqref{Eq:reg_prob}, 0.088 for the denoised estimates produced by the distributed lasso with the exact wavelet operator,
and 0.079 for the denoised estimates produced by the distributed lasso with the approximate wavelet operator with $K=15$. Note that the approximate solution does not necessarily result in a higher mean square error than the exact solution.

\begin{algorithm}[t] \label{alg3}
\caption{Distributed lasso}
%   Inputs at Node $n$: $f_n$, $\L_{n,m}~\forall m$, $\lambda_{\max}$, and \\
%   $\left\{c_{k,j}\right\}_{j=1,2,\ldots,\eta;~k=0,1,\ldots,M}$\\ [1mm]
      Inputs at node $n$: $y_n$, $\L_{n,m}~\forall m$, $\left\{\mu_{(j-1)N+n}\right\}_{j=1,2,\ldots,J+1}$, \\
      $\overline{\lambda_{\max}}$, $\gamma$, and $\left\{c_{k,j}\right\}_{j=1,2,\ldots,J+1;~k=0,1,\ldots,K}$ \\
     % and   \\
   Outputs at node $n$: $y_{n*}$, the denoised estimate of $f^0_n$ \\ %[-2mm]
       \begin{algorithmic}[1]
   \STATE Arbitrarily initialize $\left\{\left(\tilde{{a}}^{(0)}\right)_{(j-1)N+n}\right\}_{j=1,2,\ldots,J+1}$
   \STATE Set $\beta=1$
   \STATE Compute and store $\left\{\left({{\Phi}}{y}\right)_{(j-1)N+n}\right\}_{j=1,2,\ldots,J+1}$ \\ via Algorithm 1
   \WHILE{stopping criterion not satisfied}
   \STATE Compute and store 
   \begin{align*}
   \left\{\left(\tilde{{\Phi}}\tilde{{\Phi}}^*\tilde{{a}}^{(\beta-1)}\right)_{(j-1)N+n}\right\}_{j=1,2,\ldots,J+1}
   \end{align*} via Algorithm 2, followed by Algorithm 1
   \FOR{$j=1,2,\ldots,J+1$}
   \STATE Compute and store
   \begin{align*}
   &\left(\tilde{{a}}^{(\beta)}\right)_{(j-1)N+n} \\
   &=\Scal_{\left({\mu}_{(j-1)N+n}\right) {\gamma}}\left(
   \begin{array}{l}
   {\tilde{a}}^{(\beta-1)}_{(j-1)N+n}\\+{\gamma}\left(\tilde{{\Phi}}{y}\right)_{(j-1)N+n}\\
   -{\gamma}\left(\tilde{{\Phi}}\tilde{{\Phi}}^*\tilde{{a}}^{(\beta-1)}\right)_{(j-1)N+n}
   \end{array}
   \right)
   \end{align*}
   \ENDFOR
   \STATE Set $\beta=\beta+1$
   \ENDWHILE
   \FOR{$j=1,2,\ldots,J+1$}
   \STATE Set $\left(\tilde{{a}}_*\right)_{(j-1)N+n}=\left(\tilde{{a}}^{(\beta)}\right)_{(j-1)N+n}$
   \ENDFOR
   \STATE Compute and store $y_{n*}=\left(\tilde{{\Phi}}^*\tilde{{a}}_*\right)_n$ via Algorithm 2
   \STATE Output $y_{n*}$
   \end{algorithmic}
   \end{algorithm}

\section{Concluding Remarks} \label{Se:conclusion}

We presented a novel method to distribute a class of linear operators called %that can be represented as
unions of graph %Fourier
multiplier operators. The main idea is to approximate the graph %Fourier
multipliers by Chebyshev polynomials, whose recurrence relations make them readily amenable to distributed computation. % in a sensor network.
Key takeaways from the discussion and application examples include:
\begin{itemize}
\item A number of distributed signal processing tasks can be represented as distributed applications of unions of graph %Fourier
multiplier operators (and their adjoints) to signals on weighted graphs. Examples %of such tasks
    include distributed smoothing, denoising, inverse filtering, %deconvolution,
    and semi-supervised learning.
\item Graph Fourier multiplier operators are the graph analog of filter banks, as they reshape functions' frequencies through multiplication in the Fourier domain.
\item The amount of communication required to perform the distributed computations only scales with the size of the network through the number of edges of the communication graph, which is usually sparse. Therefore, the method is well suited to large-scale networks.
\item The approximate graph multiplier operators closely approximate the exact operators in practice, and for graph multiplier operators with smooth multipliers, an upper bound on the spectral norm of the difference of the approximate and exact operators decreases rapidly as we increase the Chebyshev approximation order.
\end{itemize}

\section{Appendix}

{\footnotesize

\begin{IEEEproof}[Proof of Proposition \ref{Prop:reg}]
The objective function in \eqref{Eq:reg_prob} is convex in $\mathbf{f}$. %, so d
Differentiating with respect to $\mathbf{f}$ shows that $\mathbf{f}_*$ is a solution to %we have that
\begin{eqnarray}\label{Eq:opt_eq}
\L^r \mathbf{f}_* + \frac{\tau}{2}(\mathbf{f}_*-\mathbf{y})=0
\end{eqnarray}
if and only if it is a solution to \eqref{Eq:reg_prob}.\footnote{In the case $r=1$, the optimality equation \eqref{Eq:opt_eq} corresponds to the optimality equation in \cite[Section III-A]{elmoataz} with $p=2$ in that paper.}
Rearranging~\eqref{Eq:opt_eq} gives $(\L^r + \frac{\tau}{2} I) \mathbf{f}_* = \frac{\tau}{2} \mathbf{y}$ and hence $\mathbf{f}_* = \frac{\tau}{2} (\L^r + \frac{\tau}{2} I)^{-1}\mathbf{y}$. This concludes the proof by noting that $\frac{\tau}{2} (\L^r + \frac{\tau}{2} I)^{-1}=g(\L)$, with $g(\lambda) = \frac{\tau}{2} \frac{1}{\frac{\tau}{2} + \lambda^r} = \frac{\tau}{\tau + 2\lambda^r}.$
\end{IEEEproof}

\begin{IEEEproof}[Proof of Proposition \ref{Prop:deconvolution}]
As in Proposition \ref{Prop:reg}, the objective function in \eqref{Eq:deconvolution} is convex in $\mathbf{f}$.
Differentiating it with respect to $\mathbf{f}$,
we have that~\eqref{Eq:deconvolution} is equivalent to
\begin{eqnarray}\label{Eq:opt_eq_deconv}
\L^r \mathbf{f}_* + \frac{\tau}{2}\mathbf{\Psi}^*(\mathbf{\Psi f}_*-\mathbf{y})=0.
\end{eqnarray}
Because $\L$ is symmetric, $\mathbf{\Psi} = g_{{\Psi}}(\L)$ is symmetric as well, allowing us to rearrange~\eqref{Eq:opt_eq_deconv} as
$\big( \L^r + \frac{\tau}{2} g_{{\Psi}}(\L)^2 \big) \mathbf{f}_* =\frac{\tau}{2}g_{{\Psi}}(\L) \mathbf{y}.$ In turn,
$\mathbf{f}_* = \frac{\tau}{2}\big( \L^r + \frac{\tau}{2} g_{{\Psi}}(\L)^2 \big)^{-1} g_{{\Psi}}(\L) \mathbf{y} = h(\L)\mathbf{y}$ with $h(\lambda) = \frac{\tau g_{{\Psi}}(\lambda)}{\tau g_{{\Psi}}^2(\lambda) +2\lambda^r}$.
\end{IEEEproof}

\begin{IEEEproof}[Proof of Proposition \ref{Prop:spec_bound}] From the definition of matrix functions, it follows for any $g$ defined on the eigenvalues $\lambda_0, \ldots, \lambda_{N-1}$ of $\mathbf P$ that 
\begin{align}
\opnorm{g(\mathbf P) }_2 &= \opnorm{ \text{diag}\big(g(\lambda_0), \ldots, g(\lambda_{N-1}) \big) }_2 \nonumber \\
&= \max_{\l\in\{0,1,\ldots,N-1\}} |g(\lambda_\ell)| \le \sup_{\lambda \in [0,\lambda_{\max}]} |g(\lambda)|. \label{eq:normdiag}
\end{align}

Recall that $\mathbf{\Phi} := \left[\mathbf{\Psi}_1; \mathbf{\Psi}_2; \ldots; \mathbf{\Psi}_{\eta}\right]$ and 
$\tilde{\mathbf{\Phi}} := \left[\tilde{\mathbf{\Psi}}_1; \tilde{\mathbf{\Psi}}_2; \ldots; \tilde{\mathbf{\Psi}}_{\eta}\right]$ are 
$\eta N \times N$ matrices composed of the $N \times N$ submatrices $\mathbf{\Psi}_j:=g_j(\mathbf{P})$ and 
$\tilde{\mathbf{\Psi}}_j:=p_j^K(\mathbf{P})$, respectively. Using~\eqref{eq:normdiag} and the definition~\eqref{Eq:BK_def} of $B(K)$, we obtain
\[
\opnorm{\mathbf{\Psi}_j-\tilde{\mathbf{\Psi}}_j}_2 = \opnorm{ (g_j - p_j^K )(\mathbf{P}) }_2 \le B(K)
\]
for every $j\in\{1,2,\ldots,\eta\}$. In turn, using norm inequalities for block matrices~\cite[Fact 9.10.2]{Bernstein2009}, it follows that
\[
 \opnorm{{\mathbf{\Phi}}-\tilde{\mathbf{\Phi}}}_2^2 \le
\sum_{j = 1}^\eta \opnorm{{\mathbf{\Psi}_j}-\tilde{\mathbf{\Psi}_j}}_2^2
\le \eta B(K),
\]
which completes the proof.
\end{IEEEproof}

\begin{IEEEproof}[Proof of Proposition \ref{Prop:denoising}] The solutions $\mathbf{a}_*$ to \eqref{Eq:lasso} and $\tilde{\mathbf{a}}_*$ to \eqref{Eq:approx_lasso} are not unique; however, their images $\mathbf{\Phi}^*\mathbf{a}_*$ and $\tilde{\mathbf{\Phi}}^*\tilde{\mathbf{a}}_*$ are unique.
To see this, for example for $\mathbf{\Phi}^*\mathbf{a}_*$, we can write \eqref{Eq:lasso} equivalently as
\begin{align*}
\argmin_{\mathbf{a},\mathbf{b}}&~\frac{1}{2}\norm{\mathbf{y}-\mathbf{b}}_2^2+\norm{{\mathbf{a}}}_{1,\boldsymbol{\mu}} \\
\hbox{~~~s.t. ~~}&~\mathbf{b}=\mathbf{\Phi}^*{\mathbf{a}}.
\end{align*}
Then by the strict convexity of $\norm{\cdot}_2^2$, the convexity of $\norm{\cdot}_{1,\boldsymbol{\mu}}$, and Lemma \ref{Le:s_conv} below, $\mathbf{\Phi}^*\mathbf{a}_*$ is unique.
\begin{lemma} \label{Le:s_conv}
Let $f_1:\Rbb^n \rightarrow \Rbb$ be strictly convex, $f_2:\Rbb^m \rightarrow \Rbb$ be convex, and $\mathbf{A} \in \Rbb^{n \times m}$. Then the solution $(\mathbf{x}^*,\mathbf{y}^*)$ to
\begin{align} \label{Eq:lem_con_prob}
\argmin_{\mathbf{x} \in \Rbb^n,~\mathbf{y}\in \Rbb^m} &~f_1(\mathbf{x})+f_2(\mathbf{y})  \\
\hbox{s.t.~~~~}~&~\mathbf{x}=\mathbf{Ay} \nonumber
\end{align}
is unique with respect to $\mathbf{x}^*$ (but not necessarily $\mathbf{y}^*$).
\end{lemma}
\begin{IEEEproof}[Proof of Lemma \ref{Le:s_conv}]
Let $(\mathbf{x}_1,\mathbf{y}_1)$ and $(\mathbf{x}_2,\mathbf{y}_2)$ be in the set \eqref{Eq:lem_con_prob}, and assume $\mathbf{x}_1 \neq \mathbf{x}_2$. Then
by linearity, $(\mathbf{x}_3,\mathbf{y}_3):=\frac{1}{2}(\mathbf{x}_1,\mathbf{y}_1)+\frac{1}{2}(\mathbf{x}_2,\mathbf{y}_2)$ satisfies $\mathbf{x}_3=\mathbf{Ay}_3$, and by the strict convexity of $f_1(\cdot)$ and convexity of $f_2(\cdot)$,
\begin{align*}
f_1(\mathbf{x}_3)+f_2(\mathbf{y}_3) &< \frac{1}{2} f_1(\mathbf{x}_1) +\frac{1}{2} f_1(\mathbf{x}_2)+\frac{1}{2} f_2(\mathbf{y}_1)+\frac{1}{2} f_2(\mathbf{y}_2) \\
&=~\min_{\left\{\mathbf{x} \in \Rbb^n,~\mathbf{y}\in \Rbb^m:~\mathbf{x}=\mathbf{Ay}\right\}} f_1(\mathbf{x})+f_2(\mathbf{y}),
\end{align*}
which is a contradiction. Thus, $\mathbf{x}_1=\mathbf{x}_2$.
\end{IEEEproof}

It follows from the first-order necessary and sufficient optimality equations of the lasso problem (see, e.g., \cite[Proposition 5.3(iv)]{combettes}) that for all $\mathbf{a} \in \Rbb^{N(J+1)}$, we have
\begin{align} \label{Eq:opt_cond_l_1}
\ip{\mathbf{y}-\mathbf{\Phi}^*\mathbf{a}_*}{\mathbf{\Phi}^*\mathbf{a}-\mathbf{\Phi}^*\mathbf{a}_*}+\norm{\mathbf{a}_*}_{1,\boldsymbol{\mu}} \leq \norm{\mathbf{a}}_{1,\boldsymbol{\mu}},
\end{align}
and similarly
\begin{align} \label{Eq:opt_cond_l_2}
\ip{\mathbf{y}-\tilde{\mathbf{\Phi}}^*\tilde{\mathbf{a}}_*}{\tilde{\mathbf{\Phi}}^*\mathbf{a}-\tilde{\mathbf{\Phi}}^*\tilde{\mathbf{a}}_*}+\norm{\tilde{\mathbf{a}}_*}_{1,\boldsymbol{\mu}} \leq \norm{\mathbf{a}}_{1,\boldsymbol{\mu}}.
\end{align}
Taking $\mathbf{a}=\tilde{\mathbf{a}}_*$ in \eqref{Eq:opt_cond_l_1} and $\mathbf{a}={\mathbf{a}}_*$ in \eqref{Eq:opt_cond_l_2}, summing \eqref{Eq:opt_cond_l_1} and \eqref{Eq:opt_cond_l_2}, and rearranging, we have
 \begin{align}\label{Eq:opt_cond_res}
 &\ip{\mathbf{y}-\mathbf{\Phi}^*\mathbf{a}_*}{\mathbf{\Phi}^*\tilde{\mathbf{a}}_*-\mathbf{\Phi}^*\mathbf{a}_*}+\ip{\mathbf{y}-\tilde{\mathbf{\Phi}}^*\tilde{\mathbf{a}}_*}{\tilde{\mathbf{\Phi}}^*\mathbf{a}_*-\tilde{\mathbf{\Phi}}^*\tilde{\mathbf{a}}_*} \nonumber \\
 &=\norm{\mathbf{y}-{\mathbf{\Phi}}^*{\mathbf{a}}_*}_2^2+\ip{\mathbf{y}-\mathbf{\Phi}^*\mathbf{a}_*}{\mathbf{\Phi}^*\tilde{\mathbf{a}}_*-\mathbf{y}} \nonumber \\
 &\quad +\norm{\mathbf{y}-\tilde{\mathbf{\Phi}}^*\tilde{\mathbf{a}}_*}_2^2+\ip{\mathbf{y}-\tilde{\mathbf{\Phi}}^*\tilde{\mathbf{a}}_*}{\tilde{\mathbf{\Phi}}^*\mathbf{a}_*-\mathbf{y}} \leq 0.
 \end{align}
 %Now we have
 Then
\begin{align}
&\norm{\tilde{\mathbf{\Phi}}^*\tilde{\mathbf{a}}_*-{\mathbf{\Phi}}^*{\mathbf{a}}_*}_2^2 \nonumber \\
&~=\norm{\mathbf{y}-{\mathbf{\Phi}}^*{\mathbf{a}}_*}_2^2+\norm{\mathbf{y}-\tilde{\mathbf{\Phi}}^*\tilde{\mathbf{a}}_*}_2^2 -2\ip{\mathbf{y}-{\mathbf{\Phi}}^*{\mathbf{a}}_*}{\mathbf{y}-\tilde{\mathbf{\Phi}}^*\tilde{\mathbf{a}}_*} \nonumber \\
&~\stackrel{\eqref{Eq:opt_cond_res}}\leq \ip{\mathbf{y}-{\mathbf{\Phi}}^*{\mathbf{a}}_*}{(\tilde{\mathbf{\Phi}}^*-{\mathbf{\Phi}}^*)\tilde{\mathbf{a}}_*}+\ip{\mathbf{y}-\tilde{\mathbf{\Phi}}^*\tilde{\mathbf{a}}_*}{({\mathbf{\Phi}}^*-\tilde{\mathbf{\Phi}}^*){\mathbf{a}}_*} \nonumber \\
&~\leq \norm{\mathbf{y}-{\mathbf{\Phi}}^*{\mathbf{a}}_*}_2~\opnorm{\tilde{\mathbf{\Phi}}^*-{\mathbf{\Phi}}^*}_2~\norm{\tilde{\mathbf{a}}_*}_2 \nonumber \\
&~\quad+\norm{\mathbf{y}-\tilde{\mathbf{\Phi}}^*\tilde{\mathbf{a}}_*}_2~\opnorm{{\mathbf{\Phi}}^*-\tilde{\mathbf{\Phi}}^*}_2~\norm{\mathbf{a}_*}_2 \label{Eq:cs_app} \\
&~\leq \norm{\mathbf{y}}_2~\opnorm{\tilde{\mathbf{\Phi}}-\mathbf{\Phi}}_2 \left(\norm{\tilde{\mathbf{a}}_*}_2+\norm{\mathbf{a}_*}_2\right), \label{Eq:last}
\end{align}
where \eqref{Eq:cs_app} follows from the Cauchy-Schwarz inequality, and \eqref{Eq:last} follows from
the facts that $\opnorm{A^*}_2=\opnorm{A}_2$ \cite[p.~309]{horn}, and $\norm{\mathbf{y}-{\mathbf{\Phi}}^*{\mathbf{a}}_*}_2 \leq \norm{\mathbf{y}}_2$ and
$\norm{\mathbf{y}-\tilde{\mathbf{\Phi}}^*\tilde{\mathbf{a}}_*}_2 \leq \norm{\mathbf{y}}_2$ by the optimality of $\mathbf{a}_*$ and $\tilde{\mathbf{a}}_*$, and the feasibility of
$\mathbf{a}=\mathbf{0}$. Finally, by the uniqueness of $\mathbf{\Phi}^*\mathbf{a}_*$, %from above,
$\norm{\mathbf{a}_*}_{1,\boldsymbol{\mu}}$ is the same
for all solutions $\mathbf{a}_*$, and %these are bounded as follows:
\begin{align} \label{Eq:last_one}
&\big\{\min_i \mu_i\big\} \norm{\mathbf{a}_*}_2 \nonumber \\
&\quad \leq \norm{\mathbf{a}_*}_{1,\boldsymbol{\mu}} \leq \frac{1}{2}\norm{\mathbf{y}-\mathbf{\Phi}^*\mathbf{a}_*}_2^2 + \norm{\mathbf{a}_*}_{1,\boldsymbol{\mu}} \leq \frac{1}{2}\norm{\mathbf{y}}_2^2,
\end{align}
where the last inequality again follows from feasibility of $\mathbf{a}=\mathbf{0}$. The bound in \eqref{Eq:last_one} also holds for
$\big\{\min_i \mu_i\big\} \norm{\tilde{\mathbf{a}}_*}_2$, and substituting these into \eqref{Eq:last} yields the desired result.
%, $\opnorm{\tilde{\mathbf{W}}^*-\mathbf{W}^*}_2=\opnorm{\tilde{\mathbf{W}}-\mathbf{W}}_2$.
\end{IEEEproof}

}

% -------------------------------------------------------------------------
\balance
\bibliographystyle{IEEEtran}
\bibliography{distributed}

\end{document}